\documentclass[physrev,twocolumn,superscriptaddress]{revtex4-2}
\usepackage{graphicx}% Include figure files
\usepackage{amsmath,amssymb}
\usepackage{dcolumn}% Align table columns on decimal point
\usepackage{bm}% bold math
\usepackage{color}
\usepackage{hyperref}
\hypersetup{colorlinks=true,breaklinks,linkcolor=blue,urlcolor=blue,citecolor=blue}
\usepackage{xspace}
\usepackage{multirow}
\usepackage{physics}
\usepackage[version=4]{mhchem}

\newcommand{\ceb}{\ce{CeB6}}

\begin{document}

\title{Multipolar ordering from dynamical mean field theory with application to {\ceb}}

\author{Junya Otsuki}
\affiliation{Research Institute for Interdisciplinary Science, Okayama University, Okayama 700-8530, Japan}
\author{Kazuyoshi Yoshimi}
\affiliation{Institute for Solid State Physics, University of Tokyo, Chiba 277-8581, Japan}
\author{Hiroshi Shinaoka}
\affiliation{Department of Physics, Saitama University, Saitama 338-8570, Japan}
\author{Harald O. Jeschke}
\affiliation{Research Institute for Interdisciplinary Science, Okayama University, Okayama 700-8530, Japan}

\date{\today}

\begin{abstract}
Magnetic and multipolar ordering in $f$ electron systems takes place at low temperatures of order 1--10 Kelvin.
Combinations of first-principles with many-body calculations for such low-energy properties of correlated materials are challenging problems.
We address multipolar ordering in $f$ electron systems based on the dynamical mean-field theory (DMFT) combined with density functional theory.
We derive the momentum-dependent multipolar susceptibilities and interactions 
in two ways: by solving the Bethe-Salpeter (BS) equation of the two-particle Green's function and by using a recently developed approximate strong-coupling formula.
We apply the formalism to the prototypical example of multipolar ordering in {\ceb} using the Hubbard-I solver, and demonstrate that the experimental quadrupole transition is correctly reproduced. The results by the approximate formula show good agreement with those by the BS equation.
This first-principles formalism for multipolar ordering based on DMFT has applications which are beyond the reach of the traditional RKKY formula. In particular, more itinerant electron systems including $5f$, $4d$ and $5d$ electrons can be addressed.
\end{abstract}

\maketitle

\section{Introduction}
\label{sec:introduction}

Recent progress in electronic structure calculations has been expanding the target of first-principles calculations to strongly correlated materials, in particular with the help of the DMFT~\cite{Georges1996,Kotliar2006,Imada2010}.
The DMFT combined with density functional theory (DFT+DMFT) has been employed extensively to investigate the electronic structure of transition metal oxides~\cite{Held2007,Gorelov2010,Backes2016,Ahn2021}, of iron based superconductors~\cite{Aichhorn2009,Yin2011,Ferber2012}, of rare earth compounds~\cite{Haule2010,Lu2016}, and of many other materials where electronic correlations in the form of local Coulomb repulsion or Hund's rule coupling are crucial. DFT+DMFT is often employed for the theoretical support of photoemission experiments~\cite{Ebert2011}.
There is an ongoing effort to account for the momentum dependence of the self-energy~\cite{Rohringer2018}, and the efficacy of new approaches is often demonstrated for the prime example of a moderately correlated oxide, \ce{SrVO3}. For example, DMFT has been combined with the $GW$ method~\cite{Sakuma2013,Tomczak2014}, with vertex corrections~\cite{Meng2014,Galler2017} or improved by the dynamical cluster approximation~\cite{Lee2012}. 

The description of phase transitions in correlated materials is one of the objectives in DFT+DMFT investigations~\cite{Kunes2017}.
Typical examples include metallic ferromagnetism in elemental Fe~\cite{Lichtenstein1987,Halilov1998,Lichtenstein2001,Leonov2011,Belozerov2013,Szilva2013,Nomoto2020},
isostructural transitions in elemental Ce~\cite{Pickett1981,Held2001,McMahan2003,Sakai2005,Lanata2013},
the orbital ordering in Cu fluorides~\cite{Pavarini2008,Musshoff2019}.
The intersite interactions between localized moments including multipolar degrees of freedom were formulated~\cite{Pourovskii2016,Pourovskii2019}.
Ordered magnetism has been studied within DFT+DMFT~\cite{Matsumoto2009,Haule2009,Shinaoka2015,Mandal2019,Fujiwara2022}, but such investigations are less routine than the application to paramagnetic phases. In particular, the determination of two-particle susceptibilities for comparison with inelastic neutron scattering continues to be difficult~\cite{Kunes2011,Yin2014,Strand2019,Geffroy2019}.

\begin{figure}[b]
    \centering
    \includegraphics[width=0.8\linewidth]{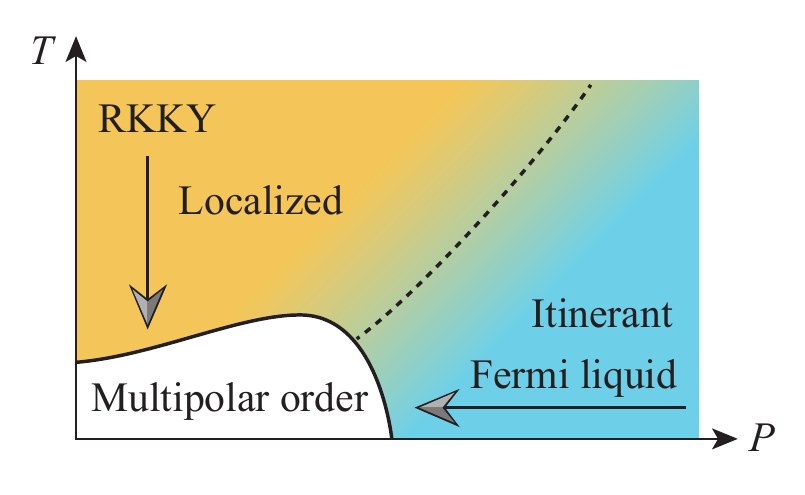}
    \caption{A schematic $P$--$T$ phase diagram of $f$ electron materials and two contrasting approaches to the multipolar ordering.}
    \label{fig:schematic}
\end{figure}

In this direction, multipolar ordering in rare-earth compounds is a challenging subject because of the $f$ orbital degrees of freedom, the strong spin-orbit coupling, and the small energy scale of order 10\,K or even less.
In this paper, we address multipolar ordering based on 
the DFT+DMFT method.
Conventionally, the multipolar ordering is considered by RKKY interactions where $f$ electrons are treated as localized.
On the other hand, a nonmagnetic Fermi liquid (heavy fermion) ground state can be addressed from the opposite limit with itinerant $f$ electrons.
Actually, the localized and itinerant $f$ states are continuously connected in a $T$-$P$ phase diagram (Fig.~\ref{fig:schematic}).
DMFT provides descriptions of the crossover between localized state and itinerant state including the heavy-fermion state near the quantum critical point~\cite{Shim2007,Otsuki2009,Otsuki2015}.
Therefore, a formalism based on DMFT covers the whole region of the phase diagram.
This motivates us to establish descriptions of multipolar ordering based on the DFT+DMFT framework for future wider applications such as mixed-valent rare-earth compounds and actinide compounds with itinerant $5f$ electrons.

\begin{figure}[t]
    \centering
    \includegraphics[width=0.9\linewidth]{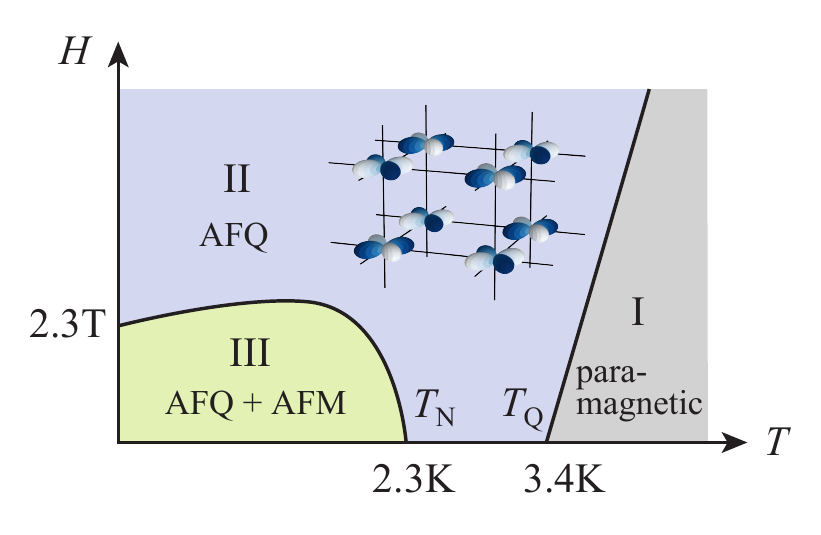}
    \caption{A schematic $T$--$H$ phase diagram of \ce{CeB6} ($\bm{H} \parallel [0, 0, 1]$)~\cite{Nakamura1995,Tayama1997}. The lattice shows the configuration of the AFQ order in phase II, where the color indicates the sign of the $xy$-type quadrupole.}
    \label{fig:CeB6}
\end{figure}

We focus on {\ceb} as a prototypical material for multipolar ordering (for a review, see Ref.~\cite{Kuramoto2009}).
The $4f^1$ configuration in trivalent Ce ions forms a
$\Gamma_8$ quartet ground state with total angular momentum $j=5/2$ under the cubic crystalline electric field (CEF)~\cite{Zirngiebl1984}.
Due to the four-fold degeneracy of the CEF multiplet, {\ceb} shows a rich phase diagram as shown in Fig.~\ref{fig:CeB6}.
Two phase transitions have been observed at low temperatures~\cite{Fujita1980}.
Below $T_\mathrm{Q}=3.4$\,K, {\ceb} exhibits the antiferro-quadrupole (AFQ) order with $\bm{q}=(1/2, 1/2, 1/2)$ (phase II)~\cite{Nakano2001,Tanaka2004,Portnichenko2020}.
The order parameter is the $(xy, yz, zx)$-type quadrupole, but the zero-field state is still under debate~\cite{Mito2023}.
At $T_\mathrm{N}=2.3$\,K, the antiferro-magnetic (AFM) order takes place on top of the AFQ order (phase III)~\cite{Effantin1985}.
Furthermore, another phase appears when Ce is partially substituted with La (phase IV)~\cite{Tayama1997}.
This phase was ascribed to the antiferro-octupole order~\cite{Kubo2004,Mannix2005,Kusunose2005}.

Theoretically, the phase transitions in {\ceb} are described by 15 multipoles in the $\Gamma_8$ quartet states~\cite{J.Ohkawa1985}.
Since the $4f$ electrons are well localized in {\ceb}~\cite{Onuki1989}, a model consisting only of the local degrees of freedom is a good starting point.
Based on a Heisenberg-type model for multipoles, Shiina \textit{et al.} derived the $T$-$H$ phase diagram and observables related to the order parameter~\cite{Shiina1997,Shiina1998}.
In this approach, the coupling constants are determined so that the model yields results consistent with experiments.

Another important issue beyond phenomenology is the microscopic derivation of the multipolar interactions to understand why the multipolar ordering takes place in individual materials.
Shiba \textit{et al.} analyzed the RKKY formula in the $\Gamma_8$ systems and deduced general properties~\cite{Shiba1999}.
Evaluation of the multipolar interactions was performed later taking explicit electronic structures into account for {\ceb}~\cite{Sakurai2004,Sakurai2005,Yamada2019,Hanzawa2019}, {\ce{NpO2}}~\cite{Kubo2005}, and Pr compounds~\cite{Iizuka2022}.

In the following, we present an application of the DFT+DMFT method to {\ceb} and reproduce the experimental order parameter.
In Sec.~\ref{sec:single-particle}, we first present the single-particle excitation spectrum, and establish how to fix parameters such as the Coulomb repulsion $U$ so that the method is applicable to other materials without any tuning parameter.
In Sec.~\ref{sec:two-particle}, the momentum-dependent multipolar susceptibilities and interactions are derived in two ways: the one using the Bethe-Salpeter equation and a recently developed approximate formula based on the decoupling of the two-particle Green's function~\cite{Otsuki2019}.
It will be demonstrated that the approximate formula can be used instead of the BS equation and shows good agreement with the experiments.

\section{Single-particle properties}
\label{sec:single-particle}

\subsection{DFT calculations}

\begin{figure}
    \centering
    \includegraphics[width=\linewidth]{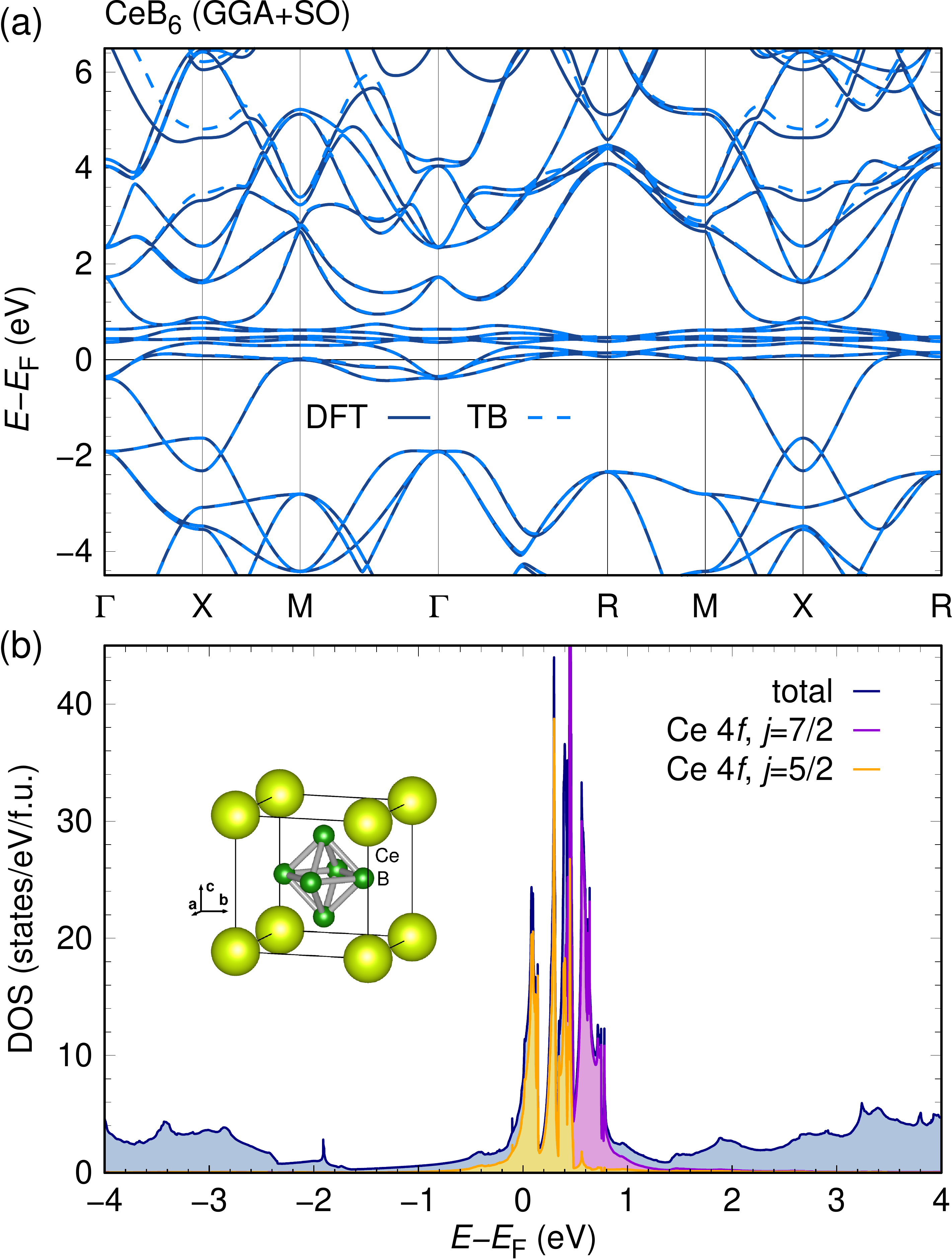}
\caption{(a) Fully relativistic band structure of \ce{CeB6} (full lines), shown together with the tight binding bands obtained from projective Wannier funcions (dashed lines). High symmetry points of the cubic $Pm\bar{3}m$ space group are $\Gamma=(0, 0, 0)$, ${\rm R}=(1/2, 1/2, 1/2)$, ${\rm M}=(1/2, 1/2, 0)$, ${\rm X}=(1/2, 0, 0)$. (b) Total and orbital-resolved density of states. The $Pm\bar{3}m$ crystal structure of \ce{CeB6} is shown as inset. }
    \label{fig:DFT}
\end{figure}

The starting point of our study is a precise DFT calculation for \ce{CeB6}. We determine its fully relativistic electronic structure using the full potential local orbital (FPLO) basis set~\cite{Koepernik1999}. The calculations are performed based on a single crystal X-ray diffraction structure determined at $T=165$\,K~\cite{Tanaka1997}. It has a cubic space group $Pm\bar{3}m$ and features one Ce atom in the corner of the cubic cell and a \ce{B6} octahedron in its center as shown in the inset of Fig.~\ref{fig:DFT}\,(b). The DFT bands are shown in Fig.~\ref{fig:DFT}\,(a) (full lines).
We use symmetry protected maximally projected Wannier functions~\cite{Eschrig2009,Koepernik2023} to construct a 72 band model consisting of Ce $4f$ and $5d$ as well as B $2s$ and $2p$ orbitals; the resulting tight binding representation of the bands is shown as dashed lines in Fig.~\ref{fig:DFT}\,(a).
Figure~\ref{fig:DFT}\,(b) shows the density of states (DOS). The Ce $4f$ partial DOS is resolved by the total angular momentum $j$ contributions.
The energy splitting $\Delta_\mathrm{SOC}$ between $j=5/2$ and $j=7/2$ due to the spin-orbit coupling (SOC) is $\Delta_\mathrm{SOC}=0.32$--0.34\,eV, which is comparable to the value in the Ce ion.
The CEF splitting between the $\Gamma_8$ quartet ground state and the excited $\Gamma_7$ doublet is 11.3\,meV ($\approx 131$\,K), which is smaller by a factor of 4 compared to the experimental value 540\,K~\cite{Zirngiebl1984}.

\subsection{DMFT calculations}
\label{sec:DMFT}

Within the DFT+DMFT method, the single-particle Green's function matrix $\hat{G}(\bm{k}, i\omega_n)$ is given by~\cite{Kotliar2006}
\begin{align}
    \hat{G}(\bm{k}, i\omega_n) = [(i\omega_n + \mu) \hat{I} - \hat{H}_\mathrm{DFT}(\bm{k}) - \hat{\Sigma}_\mathrm{loc}(i\omega_n) + \hat{\Sigma}_\mathrm{DC} ]^{-1},
\end{align}
where quantities with hat stand for $72\times 72$ matrices.
$\omega_n$ is the fermionic Matsubara frequency,
$\mu$ is the chemical potential,
and $\hat{I}$ is the identity matrix.
$\hat{H}_\mathrm{DFT}(\bm{k})$ denotes the non-interacting tight-binding Hamiltonian constructed by the DFT calculations.
$\hat{\Sigma}_\mathrm{loc}(i\omega_n)$ is the local self-energy in  DMFT, and 
$\hat{\Sigma}_\mathrm{DC}$ is the double-counting (DC) correction.

The local self-energy $\hat{\Sigma}_\mathrm{loc}(i\omega_n)$ is computed by solving the effective impurity problem for $f$ orbitals.
We treat only $j=5/2$ states, neglecting $j=7/2$ states ($j$-$j$ coupling scheme).
The influence of the neglected $j=7/2$ multiplet was analyzed in detail in Refs.~\onlinecite{Hotta2004,Hotta2006}.
We employ fully rotationally invariant Slater interactions, which are specified by four parameters called Slater integrals $F_0$, $F_2$, $F_4$, and $F_6$.
We use the standard parameterization that relates the Slater integrals to two intuitive parameters, the Coulomb repulsion $U$ and the Hund's coupling $J_\mathrm{H}$~\cite{Anisimov1997}.

We represent the DC self-energy in a simple form
\begin{align}
    \hat{\Sigma}_\mathrm{DC} = -\Delta \epsilon_f \hat{N}_f,
\end{align}
where $\hat{N}_f$ is a projection onto $f$ orbitals.
$\Delta \epsilon_f$ stands for the energy shift of the single-particle energy of the $f$ electrons.
This expression is motivated by the fact that the local electronic structure such as the SOC and the CEF are already taken into account by the DFT and do not need to be changed.
The $\hat{\Sigma}_\mathrm{DC}$ cancels only the averaged Hartree shift in $\hat{\Sigma}_\mathrm{loc}(i\omega_n)$ so that $4f$ electrons remain on the Ce atom.

We thus need to determine the three parameters $U$, $J_\mathrm{H}$, and $\Delta \epsilon_f$, and we now explain how we arrive at $U=6.2$\,eV, $J_\mathrm{H}=0.8$\,eV, and $\Delta \epsilon_f=-1.6$\,eV.
We do this based on the demand that the calculated single-particle excitation spectrum should agree with the experimental and theoretical photoemission spectroscopy (PES) and bremsstrahlung isochromat
spectroscopy (BIS) spectra (for BIS, we resort to elemental Ce).
Two energies characterize the spectrum: $\Delta_{-}$ is the minimum excitation energy from the occupied $4f^n$ multiplet to the Fermi energy $E_\mathrm{F}$, and $\Delta_{+}$ is the minimum excitation energy from $E_\mathrm{F}$ to empty $4f^{n+1}$ multiplets.
Table~\ref{tab:delta} summarizes $\Delta_{-}$ and $\Delta_{+}$ in the literature.
From the PES experiment in {\ceb}, $\Delta_{-}$ has been reported to be $\Delta_{-}=1.9$\,eV.
On the other hand, the BIS spectrum is not available for {\ceb}, and we use the data for elemental Ce instead.
The experiment reported $\Delta_{+}=3.46$\,eV, while theoretical calculation yields $\Delta_{+}=3.1$\,eV.
We adopt the theoretical value, since the theoretical result for $\Delta_{-}$ in Ce agrees with the PES value for {\ceb}.
From these observations, we adopt $\Delta_{-}=1.9$\,eV and $\Delta_{+}=3.1$\,eV. In Appendix~\ref{app:params}, we show how this leads to $U=6.2$\,eV, $J_\mathrm{H}=0.8$\,eV, and $\Delta \epsilon_f=-1.6$\,eV.

\begin{table}[]
    \centering
    \caption{Summary of the energy separation between occupied $4f^1$ state and Fermi level $\Delta_{-}$ and between occupied $4f^1$ state and lowest lying $4f^2$ state $\Delta_{+}$ in units of eV.}
    \begin{tabular}{l|cc}
        \hline
         material& $\Delta_{-}$ & $\Delta_{+}$\\
        \hline
        {\ceb} (experiment) & 1.9~\cite{Chiaia1997,Souma2001,Neupane2015} & -- \\
        Ce (experiment) & 0.27--1.92~\cite{Lang1981} & 3.46~\cite{Lang1981} \\
        Ce (theory) & 1.9~\cite{Herbst1976} & 3.1~\cite{Herbst1978} \\
        \hline
    \end{tabular}
    \label{tab:delta}
\end{table}

The DFT+DMFT cauculations were performed using open source software \texttt{DCore}~\cite{Shinaoka2021} implemented on \texttt{TRIQS}~\cite{Parcollet2015} and \texttt{DFTTools}~\cite{Aichhorn2016} libraries.
We solved the effective impurity model by exact diagonalization using \texttt{pomerol}~\cite{pomerol}.
Here, we neglected the effective hybridization $\Delta(i\omega_n)$.
This corresponds to the Hubbard-I approximation, which is often used for descriptions of properties of rare-earth compounds~\cite{Lebegue2006,Lebegue2006a,Locht2016,Delange2017}.
In fact, the Hubbard-I solver is a suitable choice for the present purpose of descriptions of the multipolar susceptibilities, in which the preservation of symmetry is relevant:
We need at least three bath sites (six states including spins) to represent the hybridization of the $j=5/2$ orbitals, which makes it difficult to compute the two-particle Green's function for the BS equation.
The de facto standard solver for DMFT, namely, the continuous-time quantum Monte Carlo method~\cite{Gull2011} approximates the interaction to density-density type to avoid a negative sign problem, which again breaks the symmetry of multipoles. Note that an approximate treatment of rotationally invariant Slater interactions was shown to improve the ferromagnetic transition temperature of elemental iron~\cite{Belozerov2013}.
Furthermore, in {\ceb}, the Fermi surface is quite similar to that of LaB$_6$~\cite{Onuki1989,Neupane2015}, indicating that the $4f$ electrons are well localized. The Kondo temperature was estimated to be of the order of 1\,K in experiments~\cite{Sato1985,Joss1987}. Therefore, a reasonable description is expected by the Hubbard-I approximation, which yields the correct N\'eel temperature for $U/W \gtrsim 1$ ($W$ is the band width) as demonstrated in Ref.~\cite{Otsuki2019}.

\begin{figure}
    \centering
    \includegraphics[width=\linewidth]{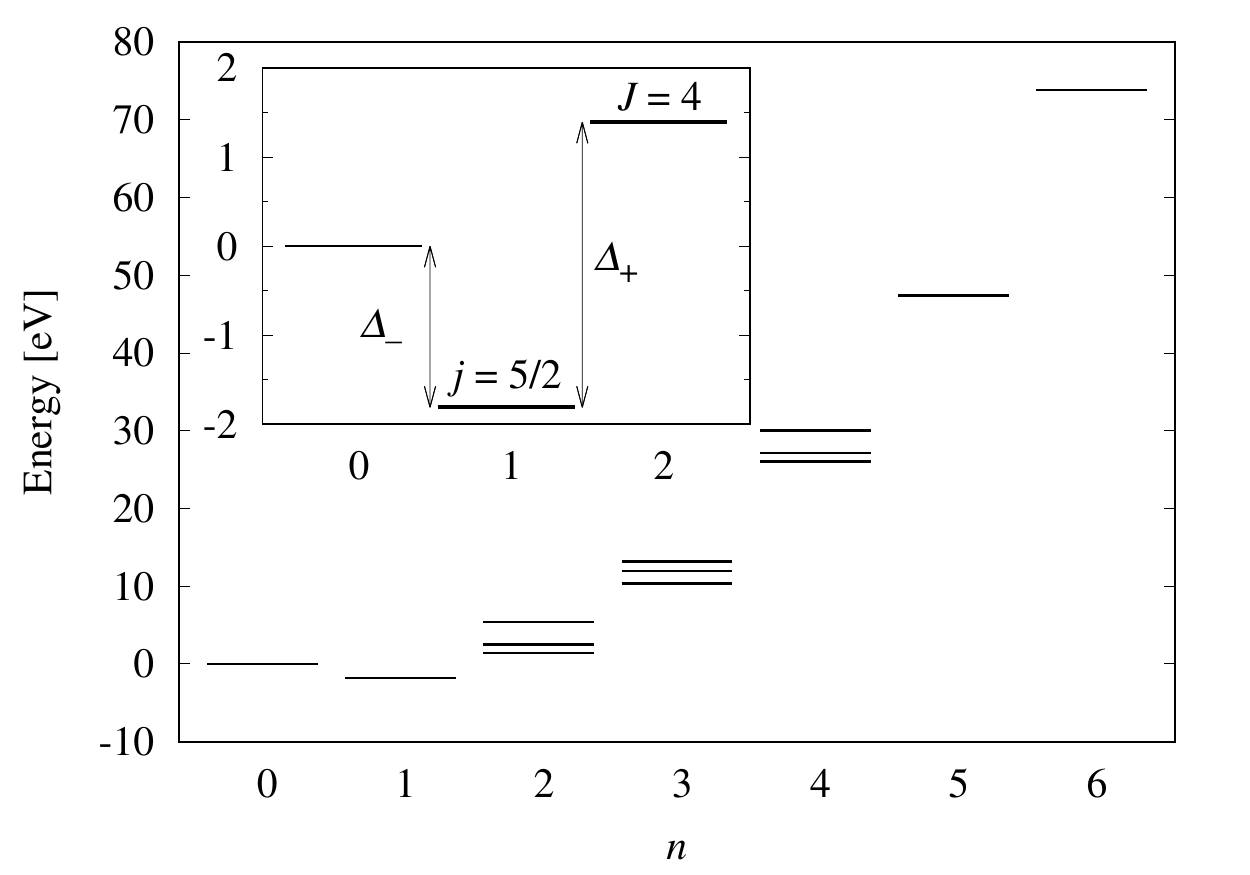}
    \caption{Eigenenergies of the many-body states of local $4f$ electrons. The inset shows a zoom up of the low-energy multiplets.}
    \label{fig:local}
\end{figure}

Without $\Delta(i\omega_n)$, energy levels of the many-body $4f$ states are well defined.
Figure~\ref{fig:local} shows the eigenvalues of the $4f^n$ multiplets
for $U=6.2$\,eV, $J_\mathrm{H}=0.8$\,eV, and $\Delta \epsilon_f=-1.6$\,eV.
The corresponding Slater integrals are 
$F_0=6.2$\,eV, 
$F_2=9.54$\,eV, 
$F_4=6.37$\,eV, and
$F_6=4.71$\,eV.
There are $2^6=64$ levels in this figure.
The CEF splitting is not visible on this scale, and we discuss only the multiplets labeled by the total angular momentum $J$.
The energy of the $j=5/2$ multiplet of the $4f^1$ configuration is $E_{1}=-1.9$\,eV ($\Delta_{-}=-E_{1}$).
The $4f^2$ configuration has three multiplets, $J=4$, 2, 0.
The Hund's ground state $J=4$ is located at $E_{2}=+1.2$\,eV, thus $\Delta_{+}=E_{2}-E_{1}=3.1$\,eV.

\begin{figure}
    \centering
    \includegraphics[width=\linewidth]{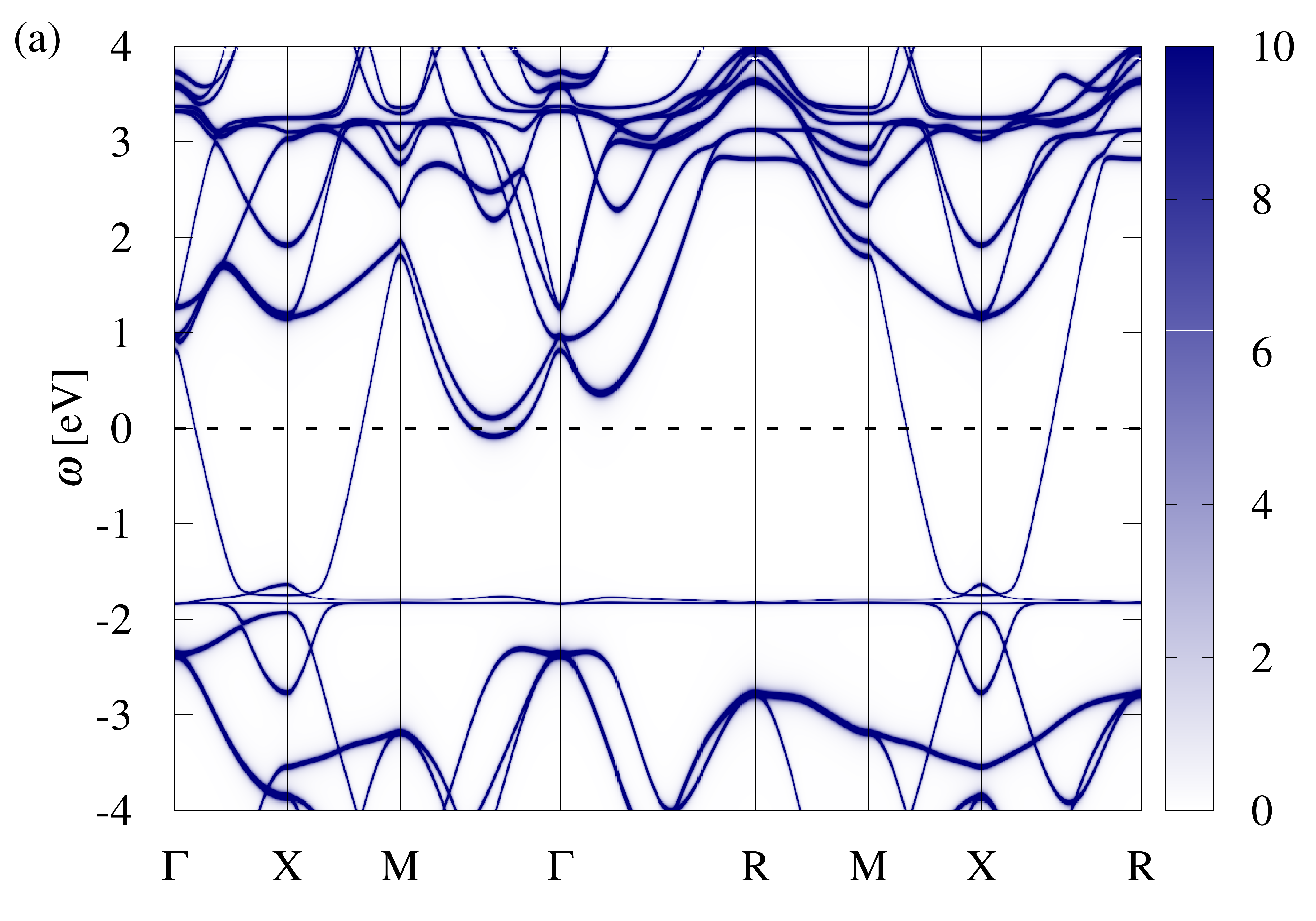}
    \includegraphics[width=\linewidth]{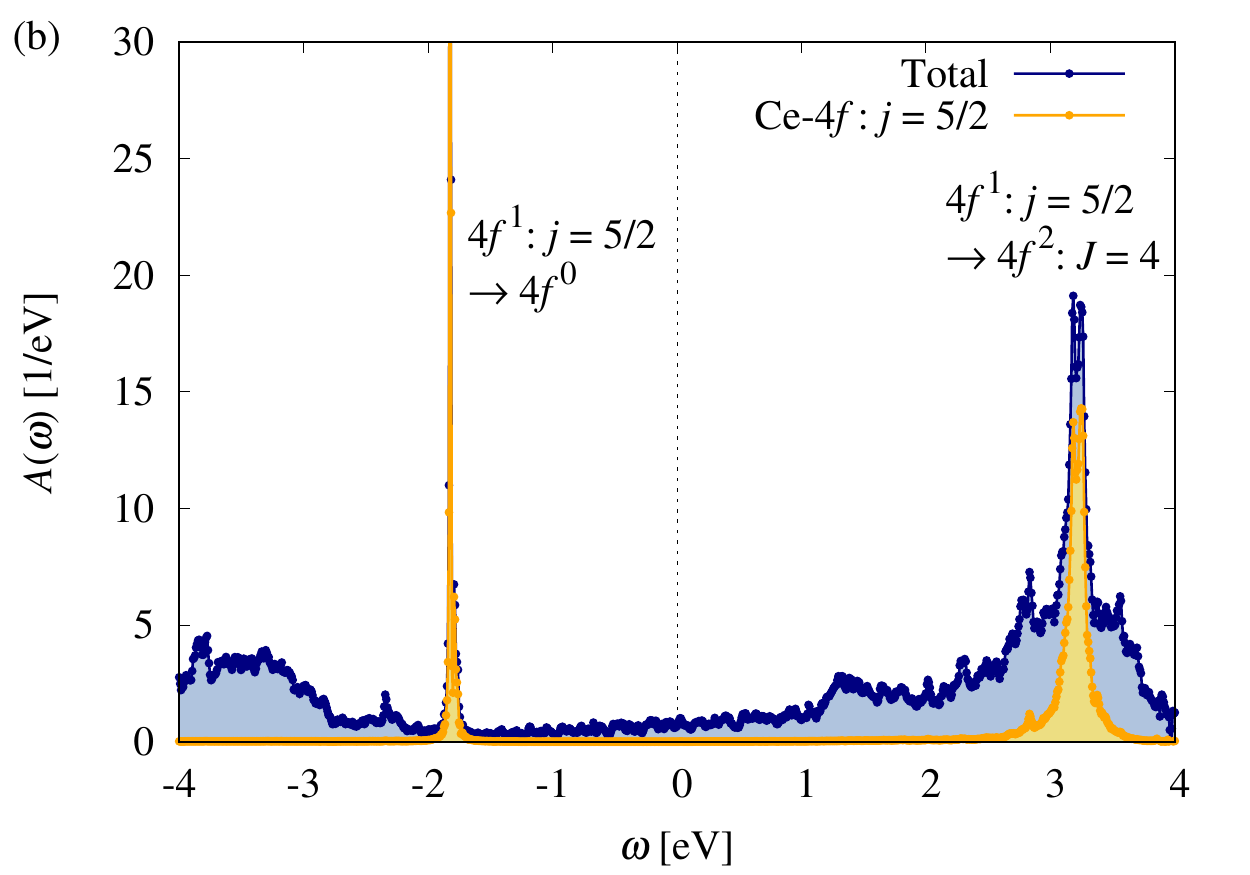}
    \caption{(a) The single-particle excitation spectrum $A(\bm{k},\omega)$ computed within DMFT. Only $j=5/2$ states are retained. A broadening of width $\delta=0.01$\,eV is introduced. (b) The $\bm{k}$-integrated spectrum $A(\omega)$ at a temperature $T=0.01$\,eV. The labels near the peaks indicate the initial and final states in the excitation process.}
    \label{fig:akw}
\end{figure}

The final results for the single-particle excitation spectrum $A(\bm{k}, \omega)$ and the $\bm{k}$-integrated spectrum $A(\omega)$ are shown in Fig.~\ref{fig:akw}.
The $4f$ peaks near $E_\mathrm{F}$ in the DFT result [Fig.~\ref{fig:DFT}\,(b)] are moved away from $E_\mathrm{F}$ in the presence of interactions. The peak at $-1.9$\,eV corresponds to electron removal from the $j=5/2$ multiplet of the $4f^1$ configuration ($4f^0$ final state), and the peak at 3.1\,eV corresponds to electron addition and the final state is the $J=4$ multiplet of the $4f^2$ configuration.
The resultant energy dispersion near $E_\mathrm{F}$ is similar to LaB$_6$~\cite{Onuki1989,Harima1988}.
A parabolic band around the X point forms a Fermi surface that is connected.
The occupied part of our spectral function, Fig.~\ref{fig:akw}\,(a), agrees very well with photoemission experiment~\cite{Koitzsch2016} and is therefore a good starting point for the determination of two-particle quantities.

\section{Two-particle properties}
\label{sec:two-particle}

\subsection{Multipolar susceptibility}

The static susceptibility describing fluctuations within $j=5/2$ states is given by
\begin{align}
    \chi_{m_1 m_2,m_3 m_4}(\bm{q}) =\int_0^{\beta}d\tau \langle O_{m_1 m_2}(\bm{q},\tau) O_{m_4 m_3}(-\bm{q}) \rangle.
\end{align}
Here, the argument $\tau$ stands for the Heisenberg operator.
The operator $O_{mm'}(\bm{q})$ is the Fourier transform of the local density operator defined by
\begin{align}
O_{mm'}(i) = f_{im}^{\dag} f_{im'},
\end{align}
where $f_{im}^{\dag}$ and $f_{im}$ are the creation and annihilation operator for $f$ electrons, respectively.
The subscript $m$ stands for the eigenvalues of $j_z$, namely, $m=-5/2$, $-3/2$, $\cdots$, $+5/2$.
Therefore, there are $6^2=36$ components in the density operators,
and $36^2=1{,}296$ components in the susceptibility.
Regarding $O_{mm'}$ as a vector, $\chi_{m_1 m_2,m_3 m_4}$ can be regarded as a $(36 \times 36)$ matrix, which we denote by $\hat{\chi}(\bm{q})$.

In DMFT, $\hat{\chi}(\bm{q})$ can be computed by solving the BS equation of the two-particle Green's function~\cite{Jarrell1992, Georges1996}.
The two-particle Green's function depends on two fermionic Matsubara frequencies in addition to four $m$ indices.
We introduced a cutoff $\omega_\mathrm{max}$ for the frequency and solved the matrix equation of size $(36N_{\omega} \times 36N_{\omega})$, where $N_{\omega}$ is the number of Matsubara frequencies below $\omega_\mathrm{max}$.
All figures presented hereafter are results computed with $N_{\omega}=160$ at $T=0.01$\,eV ($\omega_\mathrm{max}\approx 5.1$\,eV), unless otherwise specified.
For quantitative discussions, we extrapolate results to $\omega_\mathrm{max} \to \infty$ using results computed with up to $N_{\omega}=400$.

\begin{figure}
    \centering
    \includegraphics[width=\linewidth]{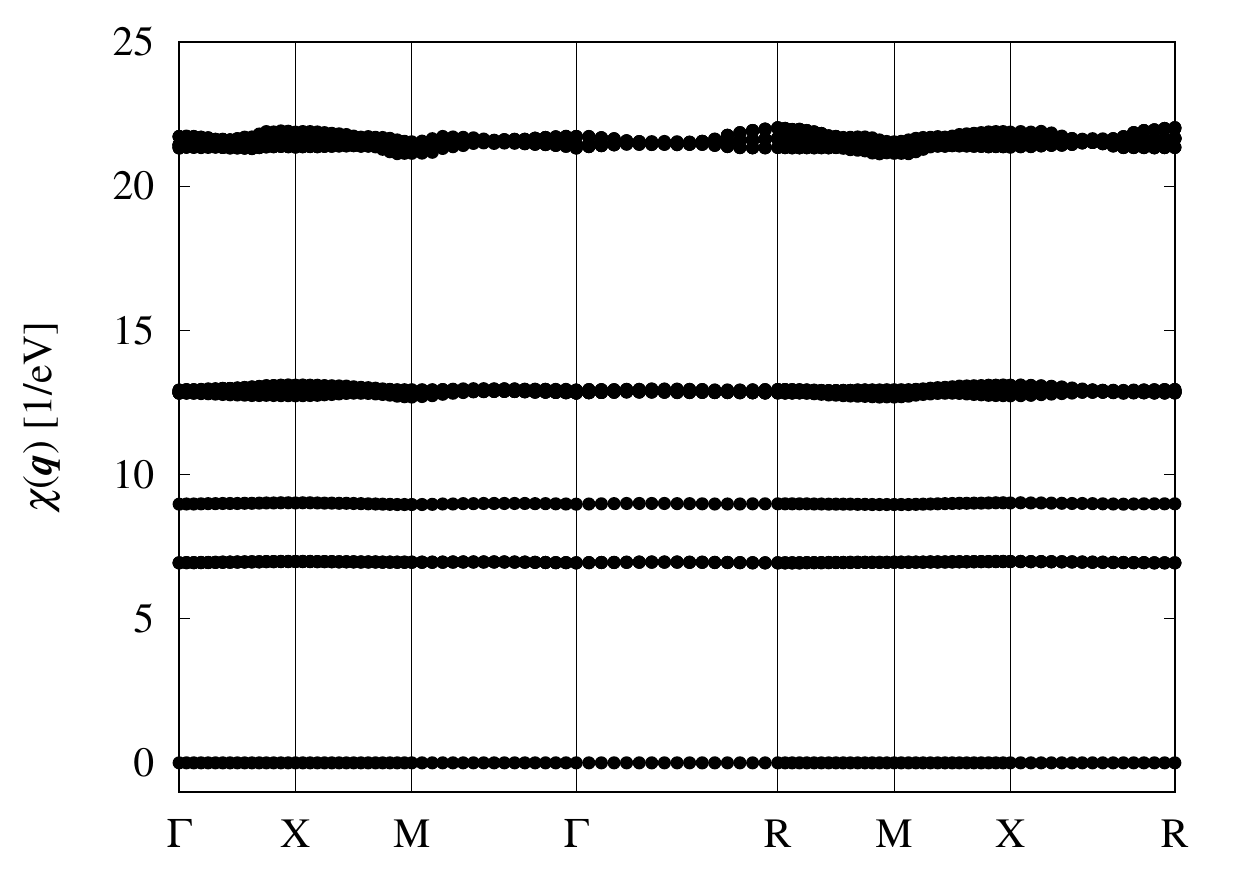}
    \caption{The $\bm{q}$-dependence of 36 eigenvalues of the susceptibility matrix $\hat{\chi}(\bm{q})$ in the $j=5/2$ states. $T=0.01$\,eV.}
    \label{fig:suscep_eigen}
\end{figure}

Figure~\ref{fig:suscep_eigen} shows the eigenvalues of the susceptibility matrix.
There are 36 eigenvalues, which are classified into 5 groups.
The leading group contains $15=4^2-1$ eigenvalues, which correspond to multipolar fluctuations within $\Gamma_8$ quartet states.
The second group contains 16 eigenvalues, which arise due to mixing between $\Gamma_8$ and $\Gamma_7$ states (two times $4\times 2$).
The third largest is the single fluctuation mode with A$_{1g}$ symmetry (hexadecapole).
The fourth largest (second smallest) eigenvalues group contains $3=2^2-1$ fluctuation modes, which correspond to the spin fluctuations within $\Gamma_7$ doublet states.
The lowest (almost zero) single-mode fluctuation is the charge fluctuation.

The eigenvectors of the susceptibility matrix yield multipoles that take the crystal symmetry into account.
Instead, we can take linear combinations of $O_{mm'}$ using pre-computed coefficients $C^{(\gamma)}_{mm'}$, namely,
\begin{align}
    O_{\gamma} = \sum_{mm'} C^{(\gamma)}_{mm'} O_{mm'},
\end{align}
where the index $\gamma$ distinguishes the irreducible representations.
In point group O$_h$, 15 operators in the $\Gamma_8$ quartet system (excluding the charge) are classified into 6 kinds of multipoles according to the rank and irreducible representations~\cite{Shiina1998,Kusunose2008,Kuramoto2009}.
Table~\ref{tab:multipoles} summarizes the basis function of these multipoles.
The corresponding expressions for $C^{(\gamma)}_{mm'}$ can be constructed from the basis set~\cite{Shiina1998,Kusunose2008,Kuramoto2009}.
We adopted a normalization $\sum_{mm'}|C^{(\gamma)}_{mm'}|^2=1$.
The multipolar susceptibilities $\chi_{\gamma}(\bm{q})$ can then be evaluated by taking linear combinations of $\chi_{m_1 m_2, m_3 m_4}(\bm{q})$ according to
\begin{align}
    \chi_{\gamma}(\bm{q}) = \sum_{m_1 m_2 m_3 m_4} C_{m_1 m_2}^{(\gamma)} \chi_{m_1 m_2, m_3 m_4}(\bm{q}) C_{m_3 m_4}^{(\gamma)\ast}.
    \label{eq:chi_gamma}
\end{align}
Other two-particle quantities having four indices are transformed into the multipolar basis $\gamma$ in a similar manner.

\begin{table}
    \centering
    \caption{The basis functions of 15 multipoles in the $\Gamma_8$ quartet system classified by the rank and irreducible representations (irreps). Extracted from Ref.~\cite{Kuramoto2009}.}
    \label{tab:multipoles}
    \begin{tabular}{ccc}
        \hline
         & Irrep & Basis \\
        \hline
        Dipole & $\Gamma_{4u}$ & $\{ x,\ y,\ z \}$ \\
        Quadrupole & $\Gamma_{3g}$ & $\{ 3z^2-r^2,\  \sqrt{3}(x^2-y^2) \}$ \\
         & $\Gamma_{5g}$ & $\{ xy,\ yz,\ zx \}$ \\
        Octupole & $\Gamma_{2u}$ & $xyz$ \\
         & $\Gamma_{4u}$ & $\{ x(5x^2-3r^2),\ y(5y^2-3r^2),\ z(5z^2-3r^2) \}$ \\
         & $\Gamma_{5u}$ & $\{ x(y^2-z^2),\ y(z^2-x^2),\ z(x^2-y^2) \}$ \\
        \hline
    \end{tabular}
\end{table}

\begin{figure}
    \centering
    \includegraphics[width=\linewidth]{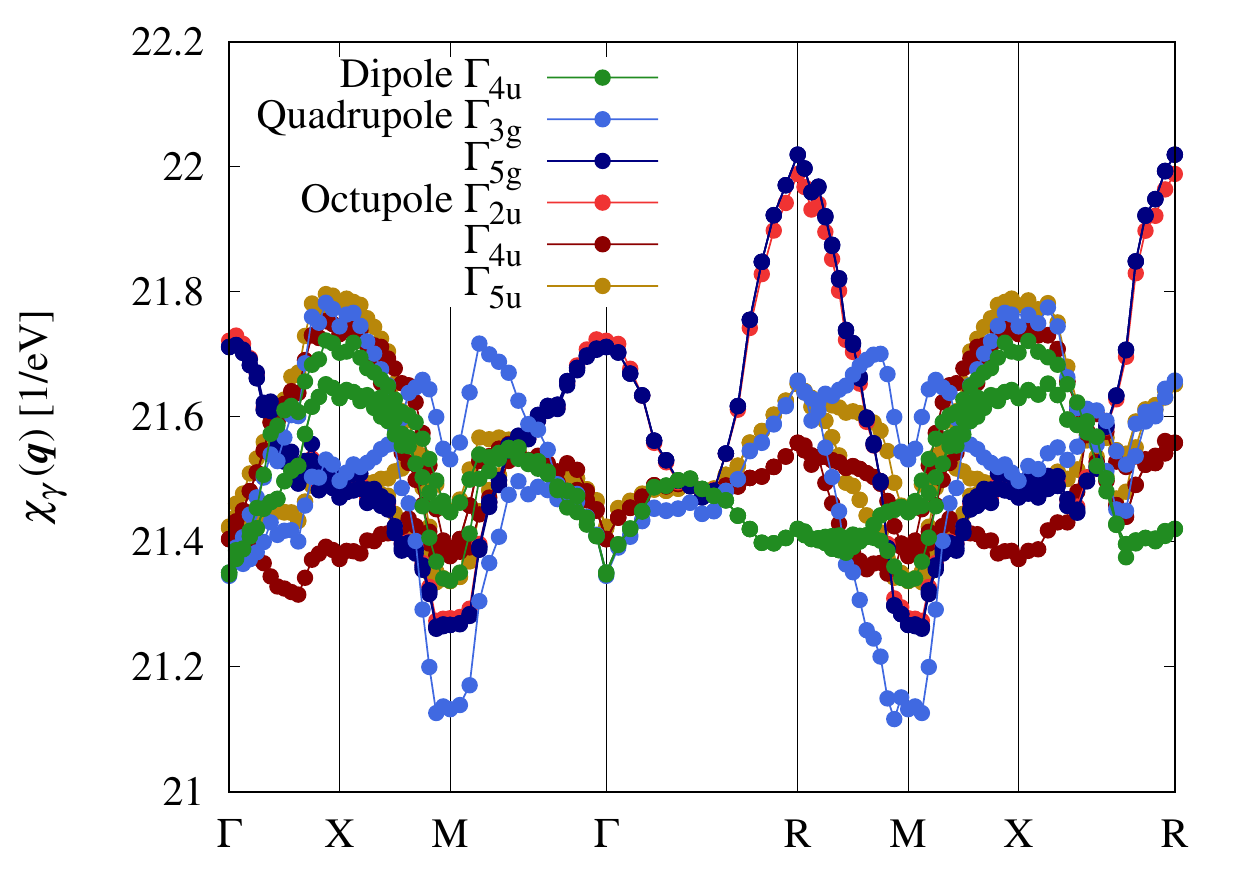}
    \caption{The $\bm{q}$ dependence of 15 multipolar susceptibilities $\chi_{\gamma}(\bm{q})$ within $\Gamma_8$ states.}
    \label{fig:suscep_Gamma8}
\end{figure}

Figure~\ref{fig:suscep_Gamma8} shows 15 susceptibilities $\chi_{\gamma}(\bm{q})$ within the $\Gamma_8$ states. 
The leading fluctuations are the quadrupoles of $\Gamma_{5g}$ symmetry at the R point [$\bm{q}=(1/2,1/2,1/2)\equiv \bm{q}_\mathrm{R}$].
This is consistent with the AFQ order of phase II in {\ceb}.
The second largest fluctuation is the octupole fluctuation of $\Gamma_{2u}$ type, which is almost degenerate with the leading quadrupolar fluctuation.
We will discuss this degeneracy later in the context of the multipolar interactions.

\begin{figure*}
    \centering
    \includegraphics[width=0.32\linewidth]{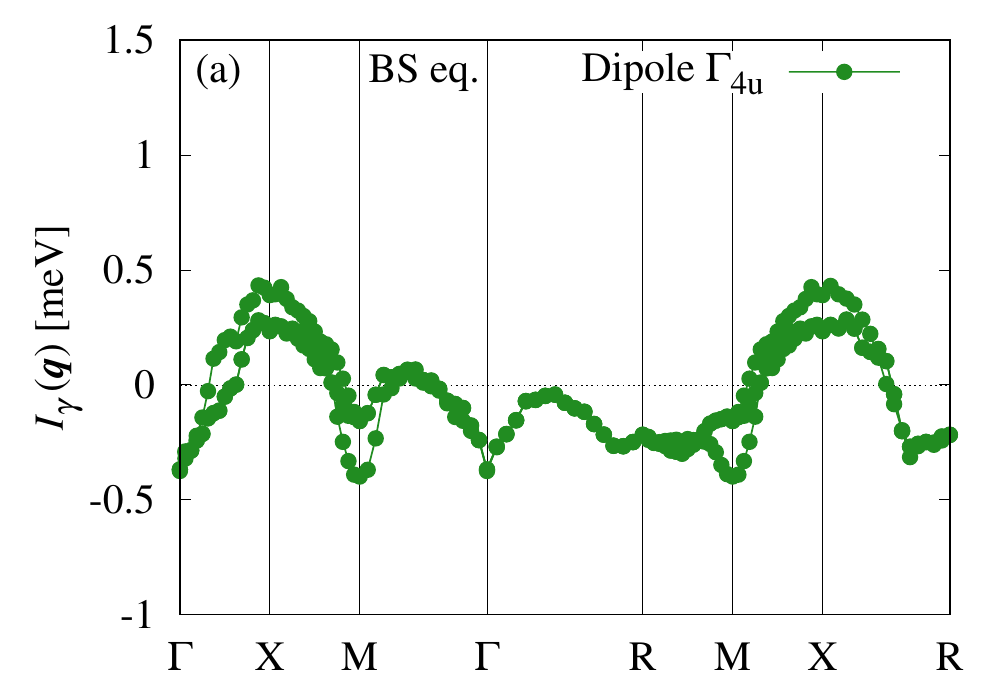}
    \includegraphics[width=0.32\linewidth]{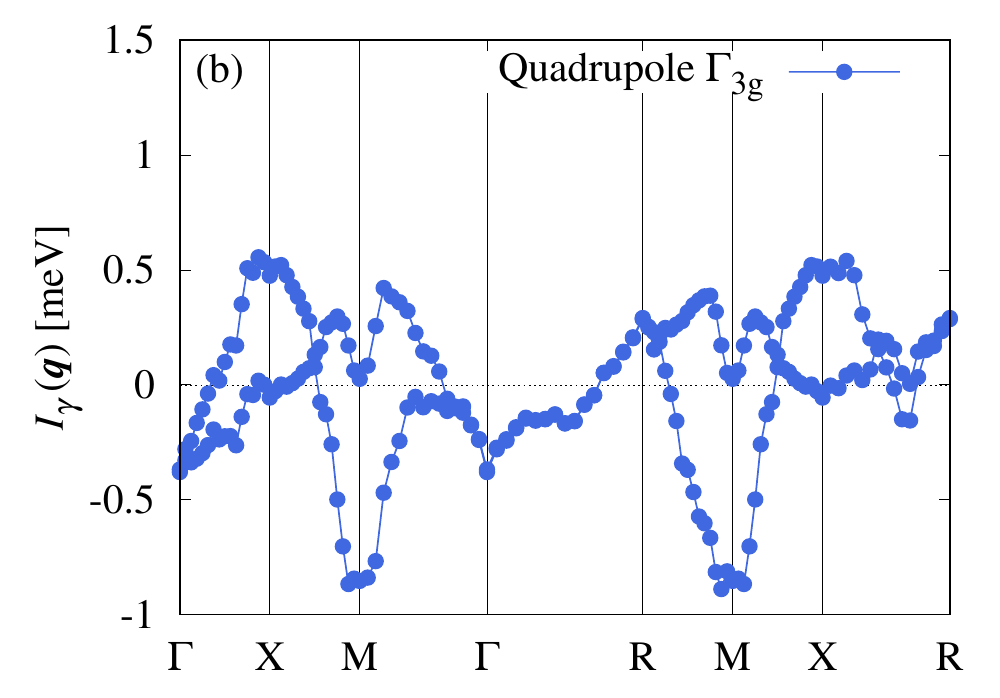}
    \includegraphics[width=0.32\linewidth]{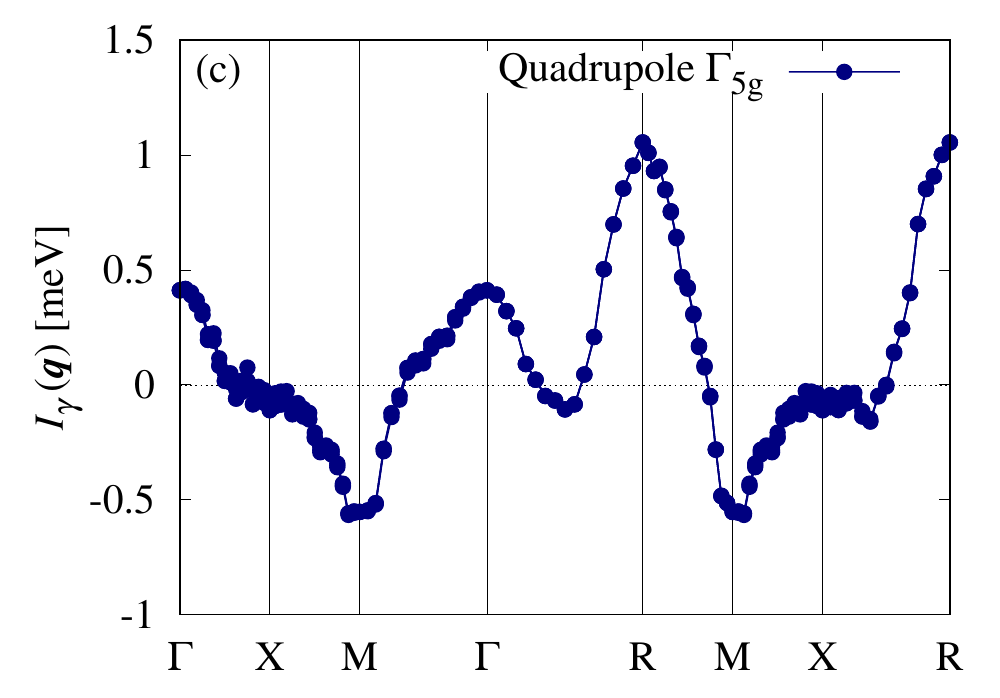}
    \includegraphics[width=0.32\linewidth]{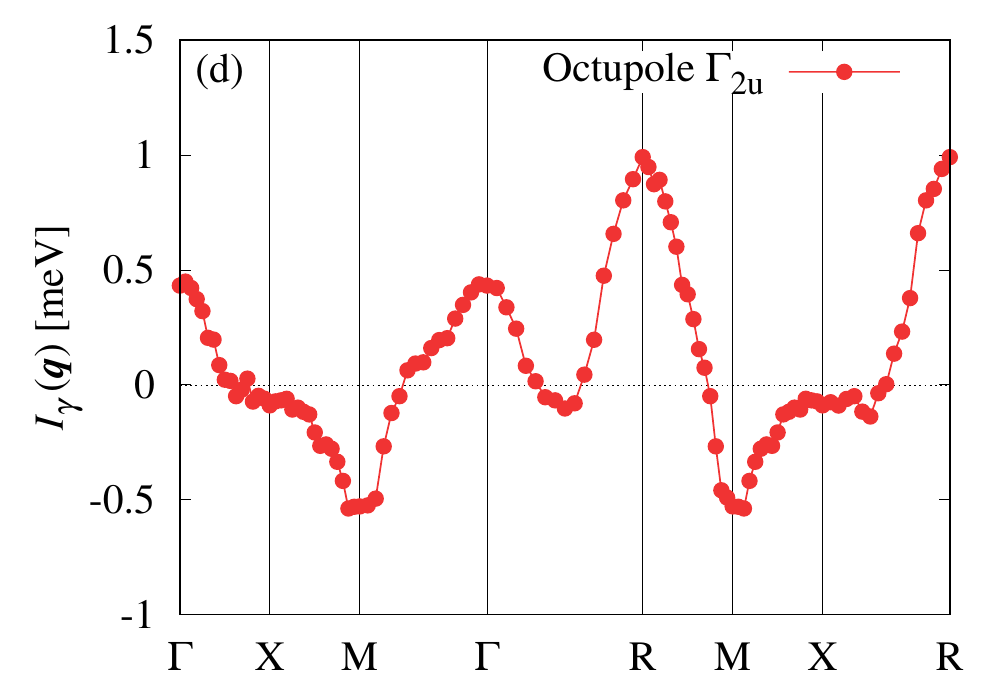}
    \includegraphics[width=0.32\linewidth]{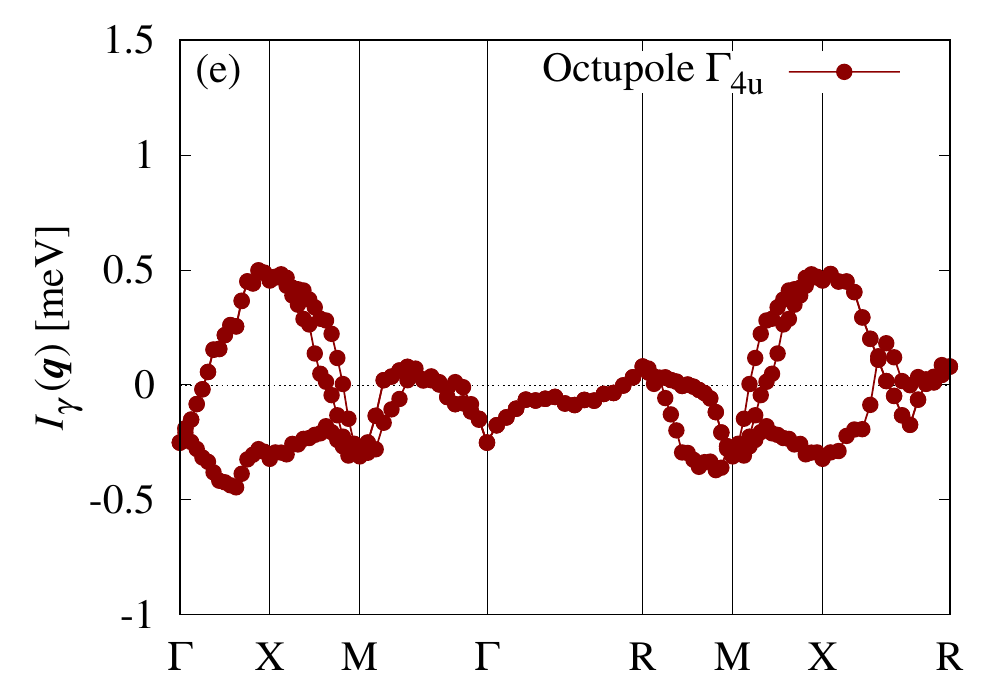}
    \includegraphics[width=0.32\linewidth]{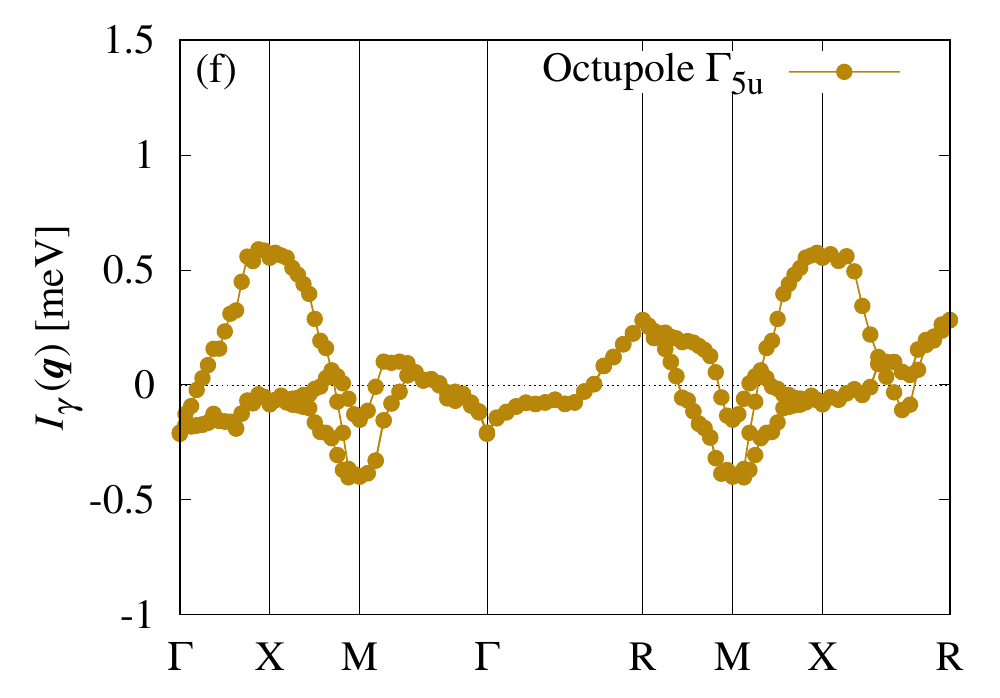}
    \caption{The $\bm{q}$ dependence of the multipolar interactions $I_{\gamma}(\bm{q})$ within $\Gamma_8$ states computed by Eq.~\eqref{eq:Iq}. Six kinds of multipoles are plotted separately. From (a) to (f) correspond to top to bottom in Table~\ref{tab:multipoles}.}
    \label{fig:Iq}
\end{figure*}

In order to determine the transition temperature, we need to decrease $T$ to detect a divergence of $\chi_{\gamma}(\bm{q})$ (see for example Ref.~\cite{Lichtenstein2001}).
However, it is practically difficult to perform the computations down to the expected transition temperature of order 1--10\,K because the required number of Matsubara frequencies, $N_{\omega}$, grows in inverse proportion to $T$.
Hence, we adopt another strategy, namely, of extracting the effective intersite interactions.
In the next two subsections, we evaluate the intersite multipolar interactions using two different approaches.

\subsection{Multipolar interactions (BS equation)}

If the $4f$ electrons are well localized, we expect that a description using an effective Heisenberg-type model with multipolar degrees of freedom is a reasonable approximation.
In this case, $\hat{\chi}(\bm{q})$ computed by solving the BS equation follows the form
$\hat{\chi}(\bm{q}) \approx [\hat{\chi}_\mathrm{loc}^{-1} - \hat{I}(\bm{q})]^{-1}$,
since DMFT treats intersite correlations at a mean-field level.
Here, $\hat{\chi}_\mathrm{loc}$ is the local susceptibility evaluated in the effective impurity model.
Assuming this expression, we \emph{define} the effective intersite interaction $\hat{I}(\bm{q})$ by
\begin{align}
    \hat{I}(\bm{q}) \equiv \hat{\chi}_\mathrm{loc}^{-1} - \hat{\chi}(\bm{q})^{-1}.
    \label{eq:Iq}
\end{align}
Once $\hat{I}(\bm{q})$ is obtained, we can extrapolate $\hat{\chi}(\bm{q})$ to lower temperatures because the dominant temperature dependence arises from $\hat{\chi}_\mathrm{loc} \propto 1/T$ in localized $4f$ electron systems.
We note that in the actual evaluation of $\hat{I}(\bm{q})$, the charge fluctuation has to be eliminated to avoid numerical instability in the matrix inversion (see Appendix~\ref{app:Iq} for details).
The multipolar interactions $I_{\gamma}(\bm{q})$ are evaluated by performing the basis transformation from $\hat{I}(\bm{q})$ as in Eq.~\eqref{eq:chi_gamma}.

Figure~\ref{fig:Iq} shows the multipolar interactions $I_{\gamma}(\bm{q})$ plotted separately for each irreducible representation.
Here, positive (negative) values enhance (suppress) the fluctuations.
The largest interaction is the quadrupole of $\Gamma_{5g}$ type at the R point [Fig.~\ref{fig:Iq}\,(c)], which induces the AFQ order in phase II.
We express this AFQ interaction by $I_\mathrm{Q}(\bm{q}_\mathrm{R})$.
Figure~\ref{fig:Iq} indicates that the interaction strength of $I_\mathrm{Q}(\bm{q}_\mathrm{R})$ is about 1\,meV.
An extrapolation to $\omega_\mathrm{max} \to \infty$ yields $I_\mathrm{Q}(\bm{q}_\mathrm{R}) \approx 1.9$\,meV (see Appendix~\ref{app:converge} for details).

\begin{figure}
    \centering
    \includegraphics[width=\linewidth]{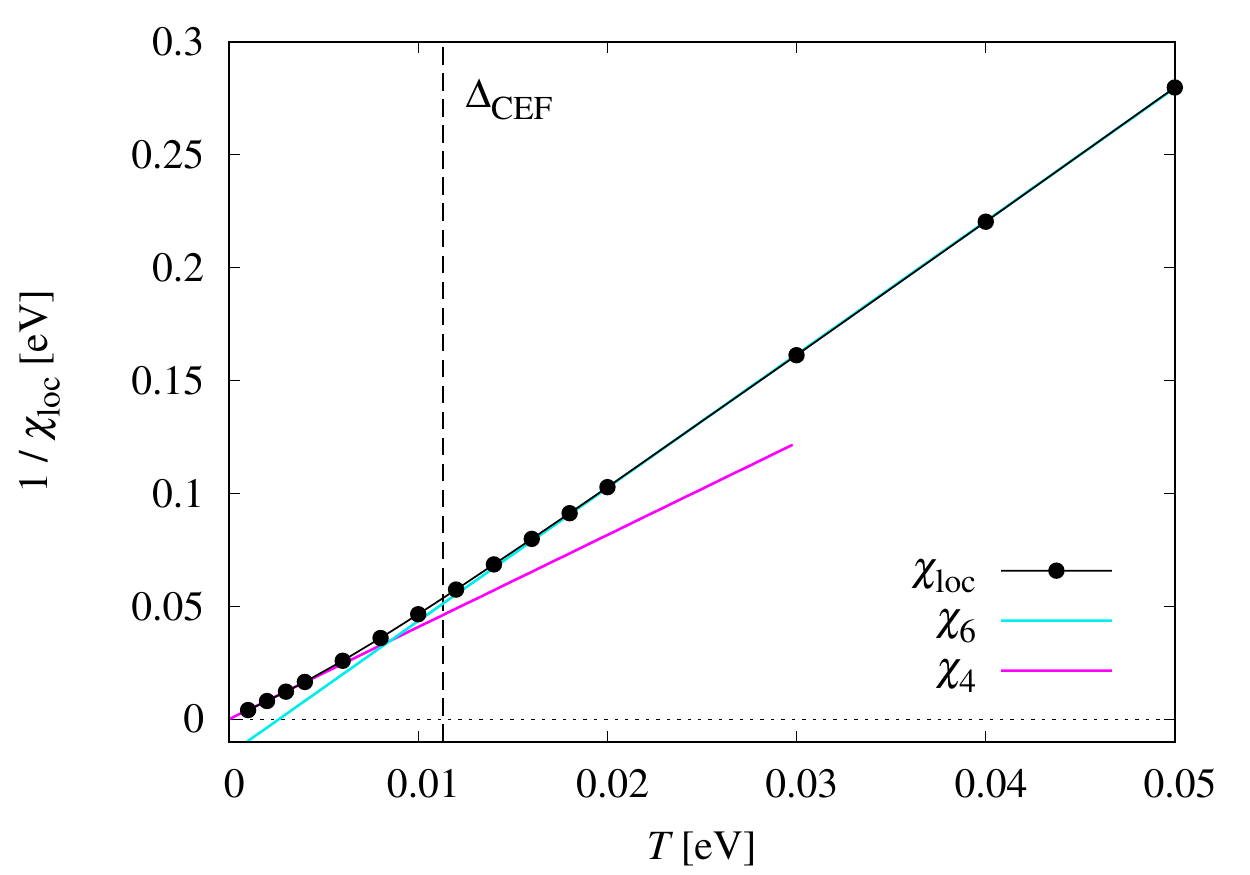}
    \caption{Temperature dependence of the local susceptibility $\chi_\mathrm{loc}$. The two lines show $\chi_6=C_6 / (T-\Theta)$ with $C_6=0.169\approx1/6$ and $\Theta=0.00316$\,eV, and $\chi_4 = C_4 / T$ with $C_4=0.245\approx1/4$. The dashed vertical line marks the energy of the CEF splitting $\Delta_\mathrm{CEF}=0.0113$\,eV.}
    \label{fig:chi_loc}
\end{figure}

The transition temperature can be determined from the divergence of the susceptibility $\hat{\chi}(\bm{q})=[\hat{\chi}_\mathrm{loc}^{-1} - \hat{I}(\bm{q})]^{-1}$. At the high-symmetry $\bm{q}$ points, the criterion of the divergence is equivalent to the condition
\begin{align}
    \chi_{\mathrm{loc},\gamma} I_{\gamma}(\bm{q}) = 1,
    \label{eq:transition_criterion}
\end{align}
where $\chi_{\mathrm{loc},\gamma}$ is the local multipolar susceptibility evaluated from $\hat{\chi}_\mathrm{loc}$ [see Eq.~\eqref{eq:chi_gamma}].
Figure~\ref{fig:chi_loc} shows the temperature dependence of $\chi_{\mathrm{loc},\gamma}$ within the Hubbard-I approximation.
Since there is no $\gamma$ dependence of $\chi_{\mathrm{loc},\gamma}$, we hereafter omit the index $\gamma$ and simply write it as $\chi_{\mathrm{loc}}$.
The two lines in Fig.~\ref{fig:chi_loc} exhibit fits by the Curie-Weiss law $\chi_6 = C_6 / (T-\Theta)$ for $T \gtrsim \Delta_\mathrm{CEF}=11.3$\,meV and the Curie law $\chi_4 = C_4 / T$ for $T \lesssim \Delta_\mathrm{CEF}$. The coefficients are obtained as $C_6=0.169\approx1/6$ and $C_4=0.245\approx1/4$ as expected from the value of the (pseudo-)degeneracy of the $4f^1$ configuration.
Using the low-temperature behavior $\chi_\mathrm{loc} \approx 1/4T$ and assuming $I_{\gamma}(\bm{q})$ to be temperature independent in Eq.~\eqref{eq:transition_criterion}, we obtain an estimate of the transition temperature for the $\Gamma_5$-type AFQ order as $T_\mathrm{Q} \approx I_\mathrm{Q}(\bm{q}_\mathrm{R})/4 \approx 0.48\,\textrm{meV} = 5.6$\,K. This is consistent with the experimental value $T_\mathrm{Q}=3.4$\,K.
The temperature dependence of $I_{\gamma}(\bm{q})$ will be discussed in Sec.~\ref{sec:SCL}.

The second largest interaction is the octupole of $\Gamma_{2u}$ type [Fig.~\ref{fig:Iq}\,(d)]. This result is consistent with the general conclusion in Ref.~\onlinecite{Shiba1999}: ``If the AFQ interaction of $\Gamma_{5g}$ type is strong, the antiferro-octupolar interaction of $\Gamma_{2u}$ type is equally strong''.
The $\Gamma_{2u}$-type antiferro-octupolar interaction plays a relevant role in stabilizing the AFQ ordering in phase II under magnetic field~\cite{Shiina1998,Kuramoto2009}.

The order parameter in phase IV is reported as the octupole of $\Gamma_{5u}$ type with $\bm{q}=\bm{q}_\mathrm{R}$~\cite{Kuramoto2009}.
Our result shows that the octupolar interaction of $\Gamma_{5u}$ type is largest at the X point and second largest at the R point.
Hence, the interaction for phase IV is not relevant in pure  \ce{CeB6} without La substitution.

\subsection{Multipolar interactions (SCL formula)}
\label{sec:SCL}

\begin{figure*}
    \centering
    \includegraphics[width=0.32\linewidth]{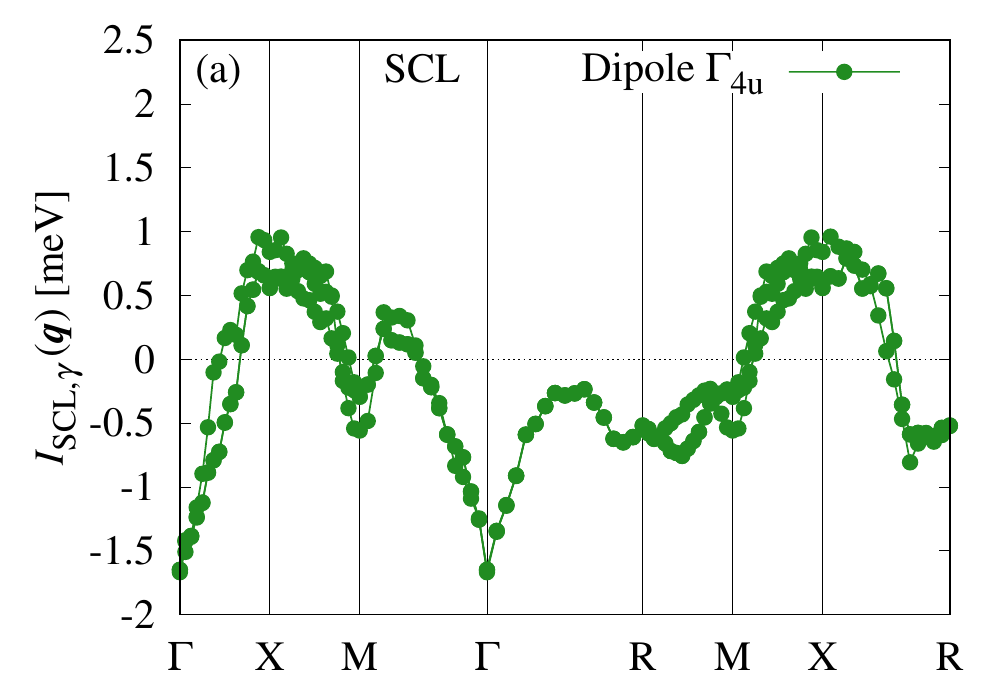}
    \includegraphics[width=0.32\linewidth]{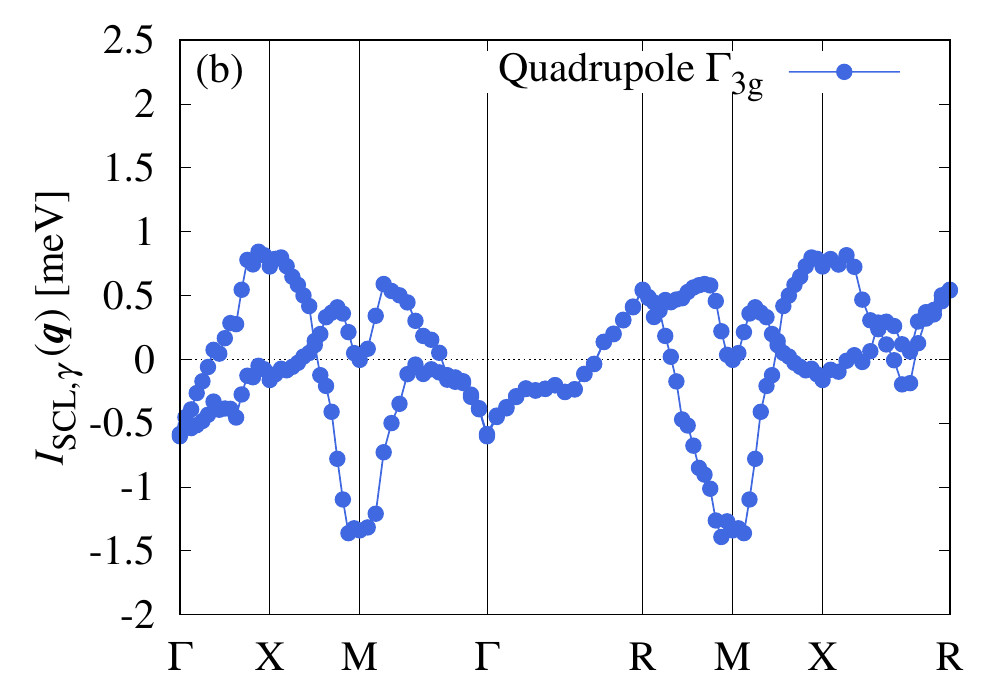}
    \includegraphics[width=0.32\linewidth]{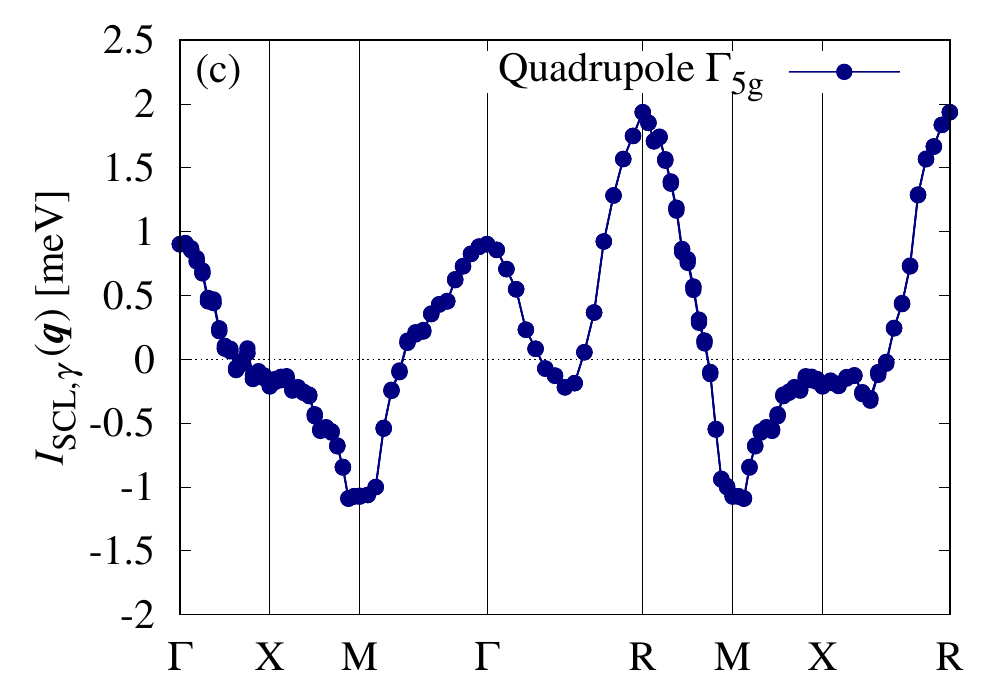}
    \includegraphics[width=0.32\linewidth]{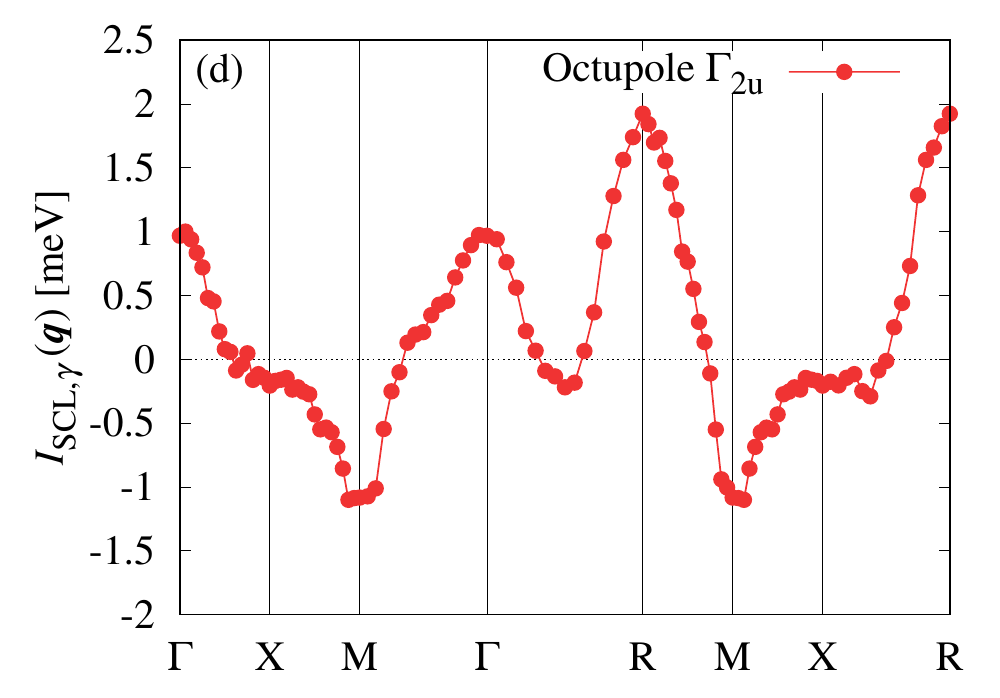}
    \includegraphics[width=0.32\linewidth]{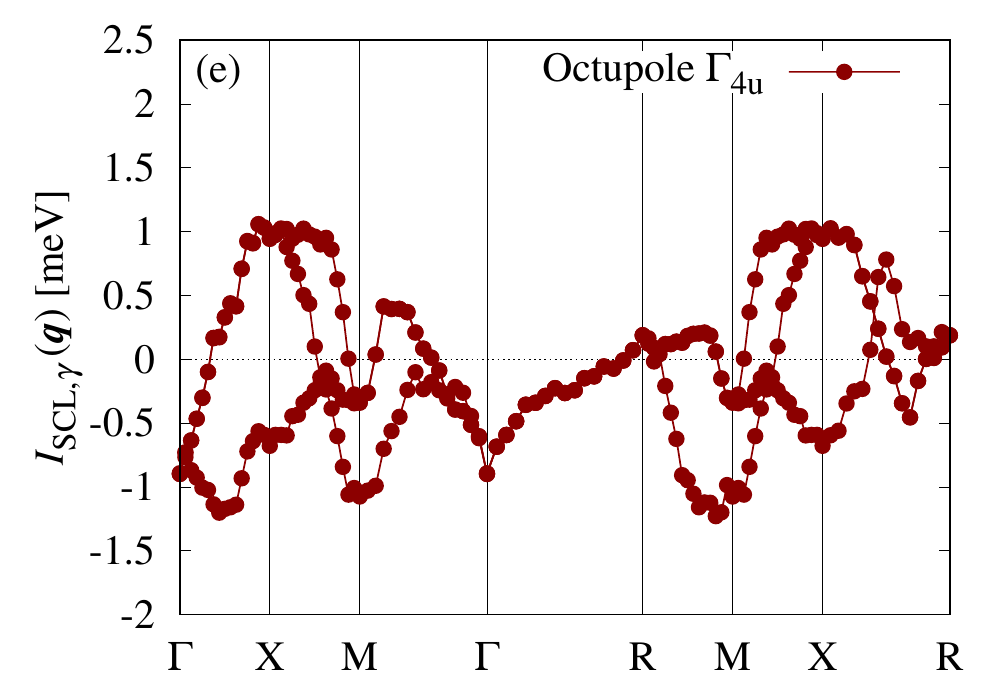}
    \includegraphics[width=0.32\linewidth]{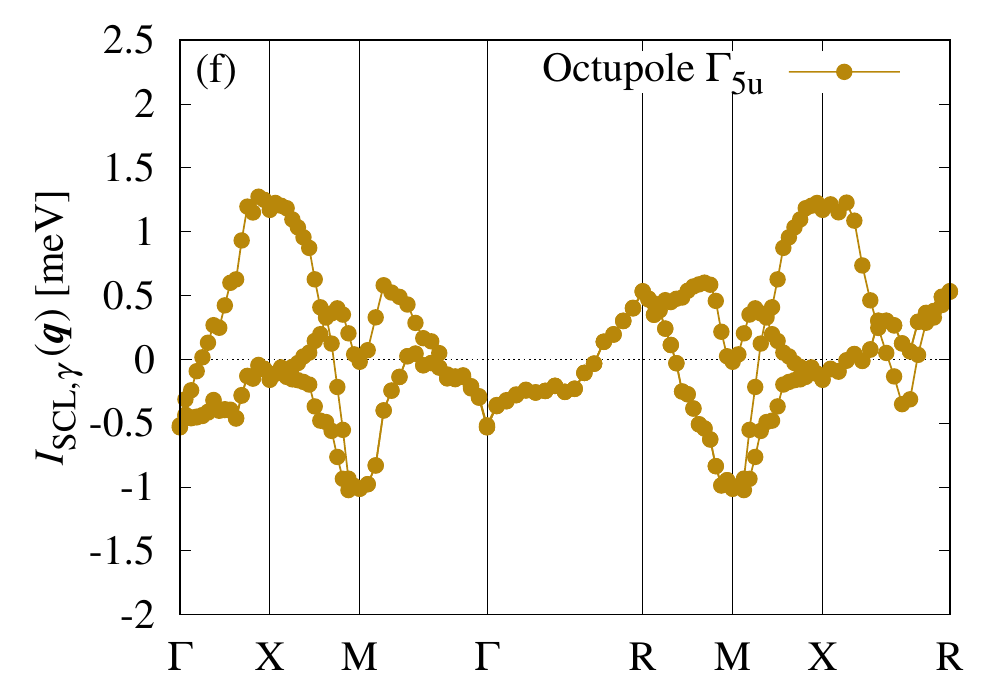}
    \caption{The $\bm{q}$ dependence of the multipolar interactions $I_{\mathrm{SCL},\gamma}(\bm{q})$ within the $\Gamma_8$ states computed by the SCL formula. See the caption of Fig.~\ref{fig:Iq} for further details.}
    \label{fig:Iq_SCL}
\end{figure*}

So far, we employed the BS equation, which requires an evaluation of the two-particle Green's function in the effective impurity model and subsequent large-scale matrix calculations.
Here, we test a recently developed approximate method named strong-coupling-limit (SCL) formula~\cite{Otsuki2019}.
In the following, we first review the SCL formula and then apply it to \ce{CeB6}.

In the SCL approximation, the momentum-dependent static susceptibility $\hat{\chi}(\bm{q})$ is given by
\begin{align}
    \hat{\chi}(\bm{q}) = [\hat{\chi}_\mathrm{loc}^{-1} - \hat{I}_\mathrm{SCL}(\bm{q})]^{-1},
\end{align}
where $\hat{I}_\mathrm{SCL}(\bm{q})$ is the intersite interaction defined by
\begin{align}
    \hat{I}_\mathrm{SCL}(\bm{q}) = T \sum_n \phi(i\omega_n)^2 \hat{\Lambda}(i\omega_n, \bm{q}).
    \label{eq:Iq_SCL}
\end{align}
The function $\hat{\Lambda}(i\omega_n, \bm{q})$ takes the energy dispersion into account and is given by
\begin{align}
    \hat{\Lambda}(i\omega_n, \bm{q}) = \hat{X}_{0, \mathrm{loc}}(i\omega_n)^{-1} - \hat{X}_{0}(i\omega_n, \bm{q})^{-1},
\end{align}
where $\hat{X}_{0, \mathrm{loc}}(i\omega_n)$ and $\hat{X}_{0}(i\omega_n, \bm{q})$ are the bare two-particle Green's functions defined by
\begin{align}
&[\hat{X}_{0}(i\omega_n, \bm{q})]_{m_1 m_2, m_3 m_4} \notag \\
&= -N^{-1}\sum_{\bm{k}} G_{m_3 m_1}(\bm{k},i\omega_n) G_{m_2 m_4}(\bm{k}+\bm{q},i\omega_n),
\end{align}
and
$\hat{X}_{0, \mathrm{loc}}(i\omega_n)=N^{-1}\sum_{\bm{q}}\hat{X}_{0}(i\omega_n, \bm{q})$
with $N$ being the number of sites.
On the other hand, the function $\phi(i\omega_n)$ takes the local fluctuations into account, which are described by the two-particle Green's function in the case of the BS equation. In Eq.~\eqref{eq:Iq_SCL}, we neglected the orbital dependence of $\phi(i\omega_n)$. In the strong-coupling regime, $\phi(i\omega_n)$ is well approximated by the two-pole function 
\begin{align}
    \phi(i\omega_n) = \frac{1}{i\omega_n + \Delta_{-}} - \frac{1}{i\omega_n - \Delta_{+}},
    \label{eq:two_pole}
\end{align}
where $\Delta_{-}$ and $\Delta_{+}$ are the excitation energies from the $4f^n$ state ($n=1$ for the Ce$^{3+}$ ion) to $4f^{n-1}$ and $4f^{n+1}$ states, respectively (see Sec.~\ref{sec:DMFT}).
The two-pole approximation corresponds to neglecting the multiplet structure in the excited $4f^{n\pm1}$ states. By using Eq.~\eqref{eq:two_pole}, we can evaluate $\hat{I}_\mathrm{SCL}(\bm{q})$ and $\hat{\chi}(\bm{q})$ without computing the two-particle Green's function and hence no extra cost of solving the effective impurity problem is incurred. This approximation was specifically termed SCL3 in Ref.~\cite{Otsuki2019}.
It has been proven analytically that $\hat{I}_\mathrm{SCL}(\bm{q})$ is reduced to the RKKY interaction or the superexchange interaction in the strong-coupling limit when it is applied to the periodic Anderson model or the Hubbard model, respectively~\cite{Otsuki2019}.
We use $\Delta_{-}=1.9$\,eV and $\Delta_{+}=3.1$\,eV as presented in Table~\ref{tab:delta}.

Figure~\ref{fig:Iq_SCL} shows the multipolar interactions $I_{\mathrm{SCL},\gamma}(\bm{q})$ computed using the SCL formula [the transformation to the multipolar basis $\gamma$ was taken as in Eq.~\eqref{eq:chi_gamma}].
It turns out that, apart from the scale of the vertical axis, the overall structure is the same as that obtained by the BS equation (Fig.~\ref{fig:Iq}). In particular, the largest interaction of the $\Gamma_{5g}$ quadrupole at the R point [Fig.~\ref{fig:Iq_SCL}\,(c)] and the second largest interaction of the $\Gamma_{2u}$ octupole [Fig.~\ref{fig:Iq_SCL}\,(d)] are correctly reproduced. 
The magnitude of the $\Gamma_{5g}$ quadrupole interaction at the R point is $I_\mathrm{Q}(\bm{q}_\mathrm{R}) \approx 1.93$\,meV, which is in surprisingly good agreement with the value obtained after extrapolation to $\omega_\mathrm{max}\to \infty$ in the BS equation.
This result demonstrates that the SCL formula is valid not only in model calculations as shown in Ref.~\cite{Otsuki2019} but also in realistic material calculations.

\begin{figure}
    \centering
    \includegraphics[width=\linewidth]{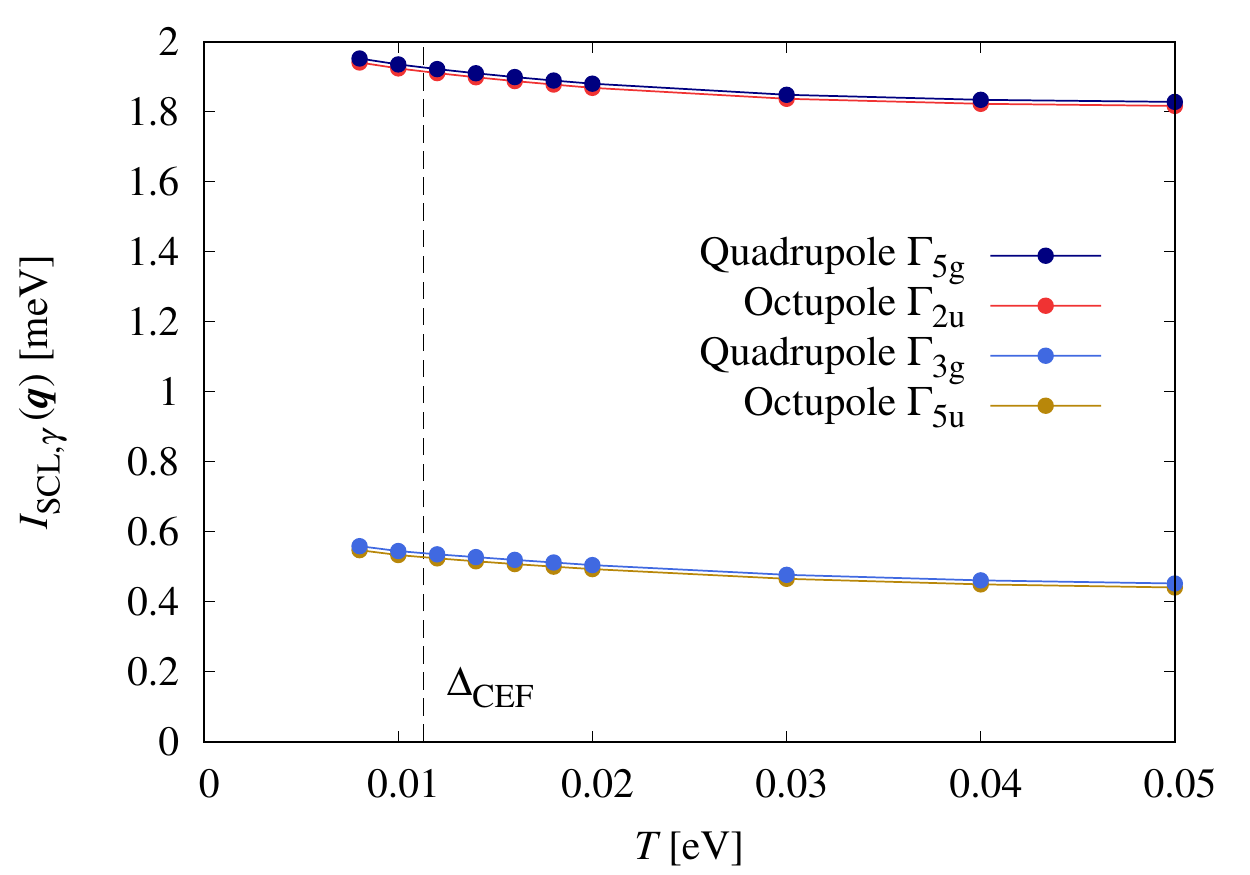}
    \caption{Temperature dependence of the multipolar interactions $I_{\mathrm{SCL},\gamma}(\bm{q}_\mathrm{R})$. The dashed vertical line marks the energy of the CEF splitting $\Delta_\mathrm{CEF}=0.0113$\,eV.}
    \label{fig:T-I_q}
\end{figure}

At the end of this section, we discuss the temperature dependence of the multipolar interactions. Figure~\ref{fig:T-I_q} shows the temperature dependence of $I_{\mathrm{SCL},\gamma}(\bm{q}_\mathrm{R})$ for multipoles $\gamma$ that have positive interaction values (that enhance fluctuations).
The results down to $T=0.008$\,eV are plotted. For $T \lesssim 0.006$\,eV, we could not take enough Matsubara frequencies to achieve convergence.
Within the computed range of temperatures, $I_{\mathrm{SCL},\gamma}(\bm{q}_\mathrm{R})$ gradually increases with decreasing temperature. This behavior is consistent with the result in the Hubbard model~\cite{Otsuki2019}.
We thus confirm that the dominant $T$ dependence in $\chi_{\gamma}(\bm{q})$ originates from $\chi_\mathrm{loc}$, which grows in proportion to $1/T$.

\section{Doping dependence}

In order to clarify which energy band plays a major role in the experimental quadrupolar ordering, we artificially shift the chemical potential and investigate how the effective interactions $I_{\gamma}(\bm{q})$ vary.
Such a ``numerical experiment'' is one of the advantages of theoretical calculations.
We shifted the chemical potential by up to $\pm1$\,eV in a rigid band treatment.
The shift of the chemical potential changes the electron number per unit cell by $-0.6$ for $\Delta\mu=-1$\,eV and $+0.9$ for $\Delta\mu=+1$\,eV, while the $4f$ electron number is unchanged.
Therefore, electron (hole) doping per boron is 0.15 (0.10) at $\Delta\mu=+1$\,eV ($-1$\,eV).
The Fermi surface changes as follows:
Under the hole doping, the open Fermi surface centered at the X point closes (the dispersion on the $\Gamma$--M line in Fig.~\ref{fig:akw} moves away from $E_\mathrm{F}$).
On the other hand, the electron doping generates small electron pockets on the $\Gamma$--M line, and hence another contribution to the susceptibility is expected.

\begin{figure}
    \centering
    \includegraphics[width=\linewidth]{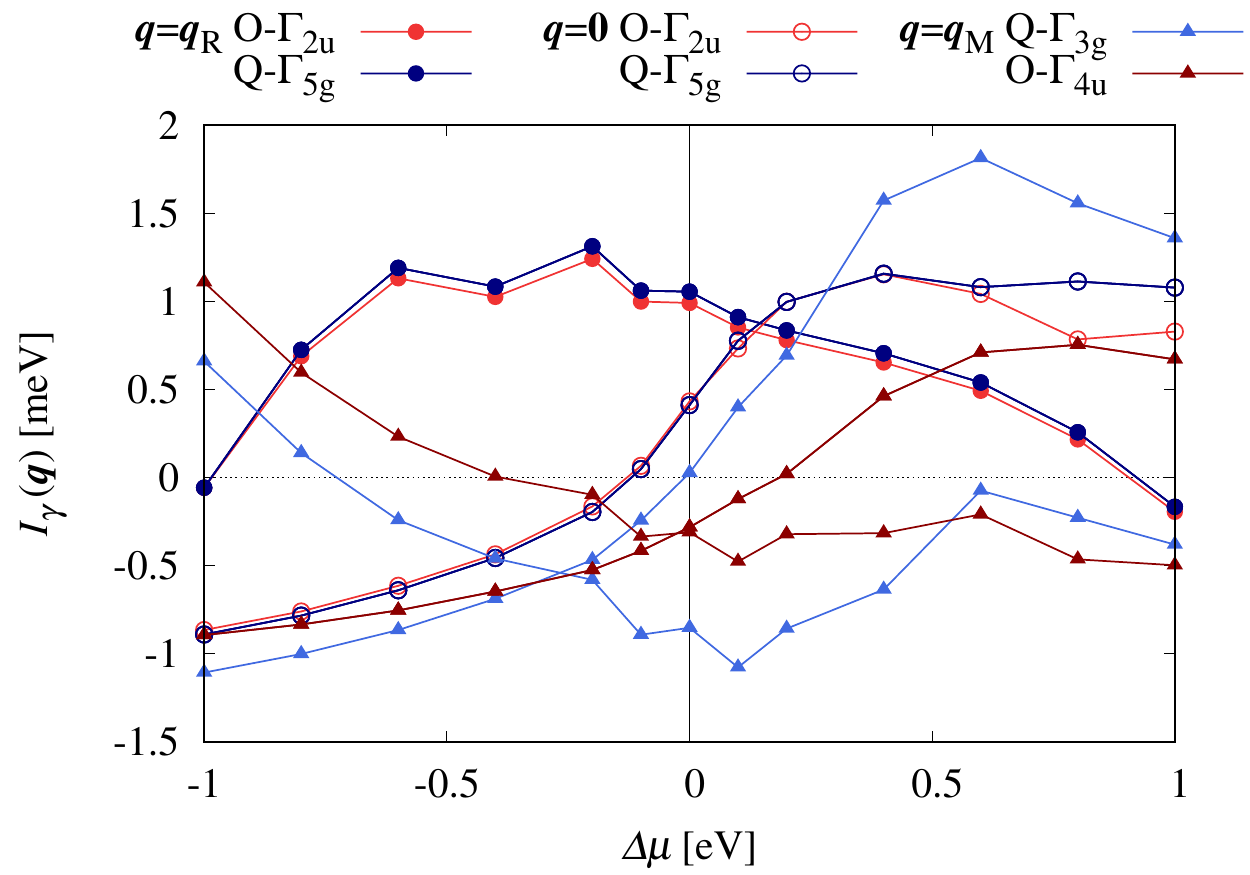}
    \caption{Variations of relevant effective interactions $I_{\gamma}(\bm{q})$ against the shift of the chemical potential, $\Delta\mu$. $I_{\gamma}(\bm{q})$ was computed by Eq.~\eqref{eq:Iq}. The symbols indicate the $\bm{q}$-vector: closed circles for the R point, open circles for the $\Gamma$ point, and closed triangles for the M point.}
    \label{fig:shift_mu}
\end{figure}

Figure~\ref{fig:shift_mu} shows the variation of the effective interactions $I_{\gamma}(\bm{q})$ as a function of $\Delta\mu$.
We plotted only relevant interactions: the $\Gamma_{5g}$-quadrupole and the $\Gamma_{2u}$-octupole at R point and $\Gamma$ point, and the $\Gamma_{3g}$-quadrupole and the $\Gamma_{4u}$-octupole at the M point.
The leading interaction $I_\mathrm{Q}(\bm{q}_\mathrm{R})$ for $\Delta\mu=0$ is enhanced by the hole doping ($\Delta\mu<0$).
Therefore, we conclude that the Fermi surface around the X point causes the experimental quadrupolar ordering.
Upon further hole doping, the octupolar interaction of $\Gamma_{4u}$ type becomes dominant.

On the other hand, $I_\mathrm{Q}(\bm{q}_\mathrm{R})$ is suppressed by electron doping ($\Delta\mu>0$), which leads to a reduction of the transition temperature of phase II.
Instead, the same multipole with a different $\bm{q}$-vector, namely, the $\Gamma_{5g}$-type ferro-quadrupolar interaction $I_\mathrm{Q}(\bm{0})$ is enhanced.
Upon further electron doping, the quadrupolar interaction of $\Gamma_{3g}$ type at the M point takes the place of the leading interaction.
Our theory thus predicts that the the quadrupolar order in phase II is robust against hole doping but is replaced by other order parameters upon electron doping.

\section{Summary}

We demonstrated a microscopic derivation of multipolar ordering using the DFT+DMFT method.
There are three parameters in these calculations: $U$, $J_\mathrm{H}$, and $\Delta \epsilon_f$.
We showed a recipe to fix them: $\epsilon_f$ is determined from the PES spectrum, and $U-J_\mathrm{H}$ can be determined from the BIS spectrum. By fixing $J_\mathrm{H}$ to a typical value, we can thus determine these parameters in an unbiased manner.

We evaluated the $\bm{q}$-dependent multipolar susceptibilities and interactions in {\ceb} by solving the Bethe-Salpeter equation and by using the SCL formula. The SCL formula, which does not require the evaluation of the two-particle Green's function, is shown to provide reasonable results that are comparable to the $\omega_\mathrm{max}\to\infty$ limit of the BS equation.
Although direct observation of the second-order phase transition is practically difficult at present, we can estimate the transition temperature by extrapolating the susceptibility to low temperatures. Our results yield a good estimate.

Comparing our approach to the more traditional combination of multiorbital Kondo lattice model and RKKY formula, there are a few advantages to the new method: (i) We base our calculations on the full DFT+DMFT spectral function so that the validity of our calculation can already be checked at the single particle level by comparing to photoemission spectroscopy experiment. (ii) Our approach is free from highly compound specific parameters that need to be determined from experiment; with only two local interaction parameters $U$ and $J_{\rm H}$ and a binding energy of the single $f$ electron, the method can be applied to numerous Ce compounds. (iii) 
Our calculation procedure based on the DFT+DMFT method is not limited to highly localized $f$ electron systems but can be also applied to itinerant $f$ electron systems.
The only difference to the calculations in this paper is the impurity solver: We have to use more sophisticated impurity solvers such as the continuous-time quantum Monte Carlo method~\cite{Gull2011} to incorporate the formation of heavy-fermion states.
Interesting applications include a hidden order in {\ce{URu2Si2}} and scalar order in {\ce{PrFe4P12}}~\cite{Kuramoto2009}.
These are two of the many possible future uses of our method.

\appendix

\begin{acknowledgments}
This work was supported by JSPS KAKENHI grants No. 18H04301 (J-Physics), No. 20K20522, No. 21H01003, No. 21H01041, and No. 23H04869.
\end{acknowledgments}

\section{Determination of parameters $U$, $J_\mathrm{H}$, and $\Delta \epsilon_f$}
\label{app:params}

\begin{figure}
    \centering
    \includegraphics[width=0.9\linewidth]{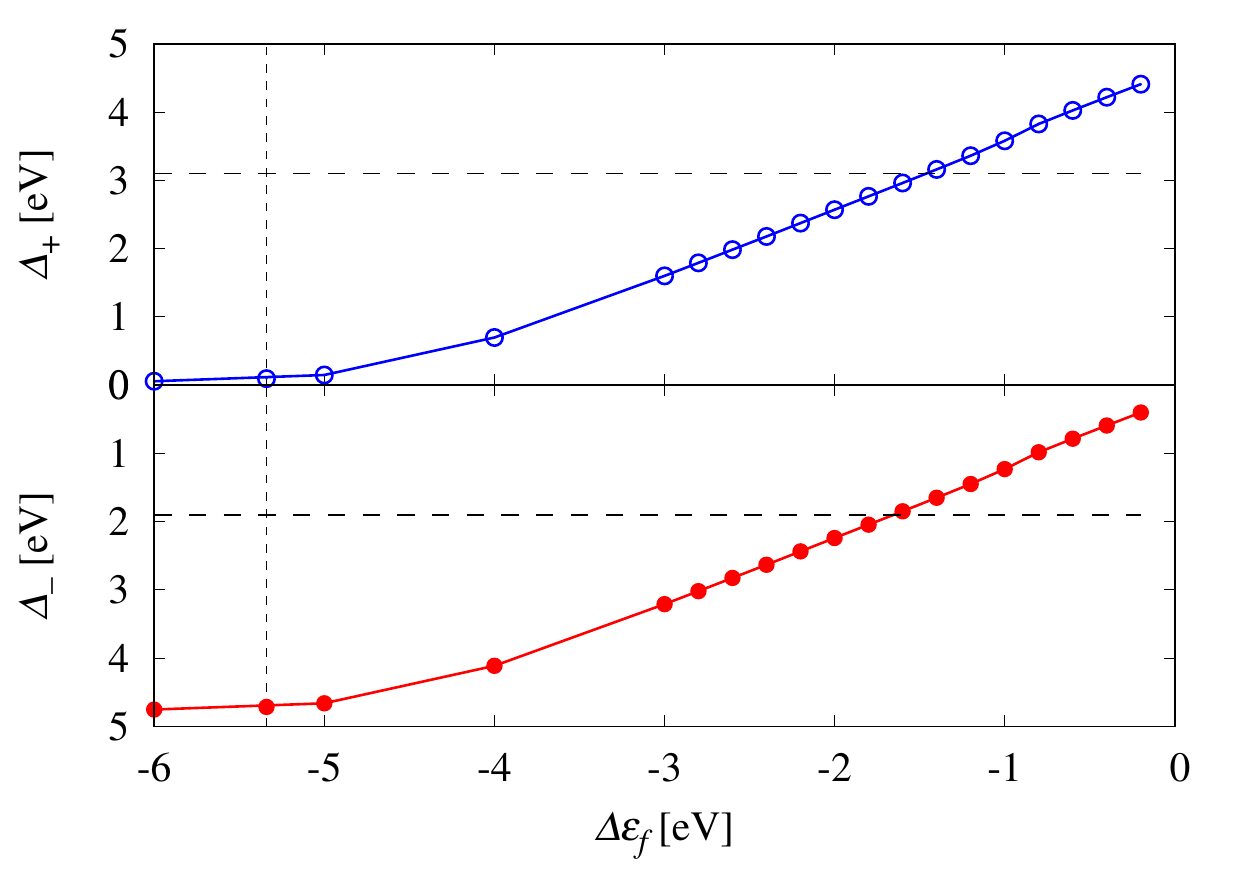}
    \caption{Calculated $\Delta_{-}$ and $\Delta_{+}$ as a function of $\Delta \epsilon_f$ for $U=6.0$\,eV and $J_\mathrm{H}=0.8$\,eV. The horizontal dashed lines indicate the target values, $\Delta_{-}=1.9$\,eV and $\Delta_{+}=3.1$\,eV.
    The vertical dashed line indicates the value of $\Delta \epsilon_f$ estimated by the Hartree energy.}
    \label{fig:ef_dep}
\end{figure}

We present the procedure of determining three parameters $U$, $J_\mathrm{H}$, and $\Delta \epsilon_f$ so that $\Delta_{-}$ and $\Delta_{+}$ agree the target values.
We first vary $\Delta \epsilon_f$ with $U$ and $J_\mathrm{H}$ fixed at 6.0\,eV and 0.8\,eV.
Figure~\ref{fig:ef_dep} shows $\Delta_{-}$ and $\Delta_{+}$ as a function of $\Delta \epsilon_f$.
The lower graph for $\Delta_{-}$ is plotted with an inverted $y$-axis because of the definition $\Delta_{-}=-E_1$. 
We find that the target value $\Delta_{-}=1.9$\,eV is obtained at $\Delta \epsilon_f=-1.6$\,eV.
For information, when we estimate $\Delta \epsilon_f$ from the Hartree energy, we obtained $\Delta \epsilon_f \approx -5.3$\,eV, which yields an unphysically deep $4f$ level.

Next, we turn our attention to $\Delta_{+}$.
The sum $\Delta_{-} + \Delta_{+}$ (the difference between $4f^1$ and $4f^2$ peaks in $A(\omega)$) depends on $U-J_\mathrm{H}$,
because $\Delta_{+}$ is the excitation energy from $E_\mathrm{F}$ to the Hund ground state of the $4f^2$ configuration. While the difference between $U$ and $J_\mathrm{H}$ is thus determined, there is still one freedom in setting $U$ or $J_\mathrm{H}$.
Following Ref.~\cite{Locht2016}, we fix $J_\mathrm{H}=0.8$\,eV and vary $U$.
Then, we found that $U=6.2$\,eV gives the target value $\Delta_{+}=3.1$\,eV.

\section{Calculation of the effective interaction}
\label{app:Iq}

Direct calculation of Eq.~(\ref{eq:Iq}) is unstable because the charge fluctuation is tiny (or zero within the computer accuracy) and the inversions of the matrices $\hat{\chi}_\mathrm{loc}$ and $\hat{\chi}(\bm{q})$ diverge.
Actually, these divergences are canceled out to yield a finite result for $\hat{I}(\bm{q})$.
We present here how to avoid this instability.

We decompose $\hat{\chi}_\mathrm{loc}$ using the singular value decomposition (SVD).
Since $\hat{\chi}_\mathrm{loc}$ is hermitian, SVD is reduced to diagonalization
\begin{align}
    \hat{\chi}_\mathrm{loc} =  U S U^{\dag},
\end{align}
where $U$ is a unitary matrix, and $S$ is a diagonal matrix whose diagonal elements (singular values) are positive and aligned in descending order.
It is clear from the above expression that small singular values make the matrix inversion unstable.
Suppose that the smallest singular value is zero, then we eliminate it and the corresponding basis from $U$.
In this singular space, $S$ and $U$ are replaced, respectively, with an $[(M-1) \times (M-1)]$ diagonal matrix $\tilde{S}$ and 
$[M \times (M-1)]$ matrix $\tilde{U}$.
Eq.~(\ref{eq:Iq}) is then expressed as
\begin{align}
    \hat{I}(\bm{q}) \simeq \tilde{U} [\tilde{S}^{-1} - (\tilde{U}^{\dag} \hat{\chi}(\bm{q}) \tilde{U})^{-1} ] \tilde{U}^{\dag}.
\end{align}
This expression does not suffer from numerical instability.

\section{Extrapolation of the frequency cutoff}
\label{app:converge}

\begin{figure}
    \centering
    \includegraphics[width=0.9\linewidth]{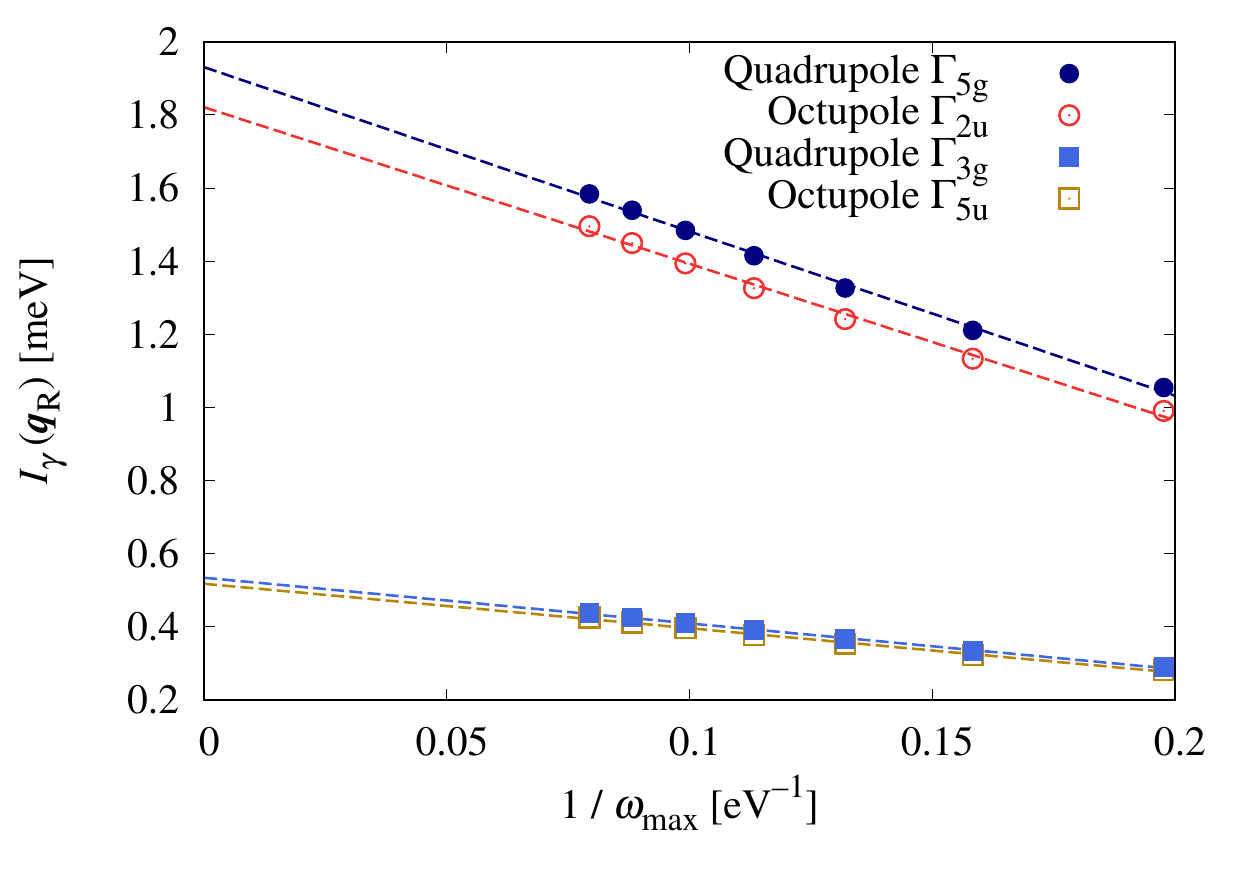}
    \caption{Extrapolation to $\omega_\mathrm{max}\to \infty$ ($1/\omega_\mathrm{max}\to 0$) for the four leading effective interactions $I_{\gamma}(\bm{q}_\mathrm{R})$ computed by the BS equation. The dashed lines show the result of linear fitting.}
    \label{fig:converge}
\end{figure}

We solved the BS equation for $\chi({\bm{q}})$ by introducing the cutoff $\omega_\mathrm{max}$ for the fermionic Matsubara frequency.
In this Appendix, we present extrapolation to $\omega_\mathrm{max} \to \infty$.
Figure~\ref{fig:converge} shows the effective interaction $I_{\gamma}(\bm{q}_\mathrm{R})$ as a function of $1/\omega_\mathrm{max}$.
Here, the results for $N_{\omega}=160,\ 200,\ \cdots,\ 400$ are plotted.
The data are well approximated by a line as indicated by the dashed lines.
Thus, extrapolation to $1/\omega_\mathrm{max} \to 0$ yields\textbf{} $I_{\gamma}(\bm{q}_\mathrm{R}) \approx 1.93$\,meV for the $\Gamma_{5g}$-type quadrupolar interaction.

More elaborate methods for solving the BS equation have been proposed~\cite{Kunes2011,Tagliavini2018,Shinaoka2020-rp,Wallerberger2021-kv,Kitatani2022-xh}.
These approaches make the convergence faster or may allow us to obtain converged results without extrapolation.

\bibliography{ceb6_paper_abbrev,manual}

\end{document}